\documentclass[11pt,fleqn,twoside]{article}

\newif\ifreport
\reporttrue

\usepackage[german,english]{babel}
\usepackage{derireport}

\usepackage{times}
\usepackage{alltt}
\usepackage{epsfig}
\usepackage{url}
\usepackage{xspace}
\usepackage{latexsym}
\usepackage{rotating} 
\usepackage{subfigure}

\usepackage{verbatim}
\usepackage[tbtags]{amsmath}
\usepackage{amsfonts}
\usepackage{amssymb}
\usepackage{cite}
\usepackage{colortbl}
\usepackage{tabularx}
\usepackage[T1]{fontenc}
\usepackage{ae,aecompl}
\usepackage{pslatex}
\usepackage{mathtools}
\usepackage{subfig}
\usepackage{pstricks}
\usepackage{fancybox}
\usepackage{ids}
\usepackage[leftcaption]{sidecap}

\newcommand{\changefont}[3]{
\fontfamily{#1} \fontseries{#2} \fontshape{#3} \selectfont}

\newcounter{myDefCounter}

\newenvironment{myDefinition}[1]
{\medskip\noindent\refstepcounter{myDefCounter}{ \textit{Definition \arabic{myDefCounter}} (\textit{#1}).}\ }
{\hspace*{\fill}~\mbox{\rule[0pt]{1.3ex}{1.3ex}}\par\endtrivlist\unskip\medskip}

\newcounter{myExampleCounter}
\newenvironment{myExample}
{
      \refstepcounter{myExampleCounter}
      \vspace{0.2cm}
      \newline
      { \changefont{ptm}{m}{sc} Example \arabic{myExampleCounter}:\ }
      \normalfont
}
{
      \hspace*{\fill}~\mbox{\rule[0pt]{1.3ex}{1.3ex}}\par\endtrivlist\unskip\vspace{0.3cm}
}

\foreignreport

\title{Finding Information Through Integrated Ad-Hoc Socializing in the
  Virtual and Physical World}

\authornames{
Christian von der Weth
\and
Manfred Hauswirth
}

\innertitle{Finding Information Through Integrated Ad-Hoc Socializing in the
  Virtual and Physical World}

\author{
Christian von der Weth%
\affiliation{
DERI (Digital Enterprise Research Institute), National University of Ireland,
Galway\protect, IDA Business Park, Lower Dangan, Galway, Ireland.
\mbox{E-mail: christian.vonderweth@deri.org.}}
\and
Manfred Hauswirth%
\affiliation{
DERI (Digital Enterprise Research Institute), National University of Ireland,
Galway\protect, IDA Business Park, Lower Dangan, Galway, Ireland.
\mbox{E-mail: manfred.hauswirth@deri.org.}}
}

\published{
}

\acknowledgement{
This work has been funded (in a part) by
    Science Foundation Ireland (Grant no.~SFI/08/CE/I1380 -- L\'{i}on-2)
}

\abstract{
Despite the services of sophisticated search engines like Google, there are a number of interesting information sources which are useful but largely inaccessible to current Web users. These information sources are often ad-hoc, location-specific and only useful for users over short periods of time, or relate to tacit knowledge of users or implicit knowledge in crowds. The solution presented in this paper addresses these problems by introducing an integrated concept of ``location'' and ``presence'' across the physical and virtual worlds enabling ad-hoc socializing of users interested in, or looking for similar information. While the definition of presence in the physical world is straightforward -- through a spatial location and vicinity at a certain point in time -- their definitions in the virtual world are neither obvious nor trivial. Based on a detailed analysis we provide an integrated spatial model spanning both worlds which enables us to define presence of users in a unified way. This integrated model 
allows us to enable ad-hoc socializing of users browsing the Web with users in the physical world specific to their joint information needs and allows us to unlock the untapped information sources mentioned above. We describe a  proof-of-concept implementation of our model and provide an empirical analysis based on real-world experiments.
\\*[\parskip]
~
\\*[\parskip]
}

\deritr{2013-03-20}
\date{March 2013}

\begin{document}

%
%

\maketitle

\newpage 
\pagenumbering{Roman}

{\small

\tableofcontents
}


\newpage
\pagenumbering{arabic}

\section{Introduction}

Support for finding information on the Internet has reached a high level of maturity. For example, Google not only indexes all Web information but also provides good results for information in online fora, social networks, and mailing lists. Despite these advances, we argue that there a various kinds of information needs that cannot be satisfied by traditional browsing or by applying existing technologies, for example: 

\textit{(1) In-situ search.} Information needs of users often refer to both a specific location and a specific, often short-term, time span, e.g., ``Is the advertised bargain still available at Shop X?'' Answers to such questions are too specific and the time frame for their validity is too short to be maintained on a web page, let alone being indexed by a search engine.  The best, and often the only possible answers can be provided by people on-site.

\textit{(2) Opinion search.} While recommender systems are integral part of many online platforms, they are typically static with no inherent support of synchronous communication between users. Writing down, e.g., a hotel review covering all relevant facets about its quality en bloc is a non-trivial task. However, such knowledge may naturally surface in a social environment, e.g., in a discussion between like-minded people.

\textit{(3) Crowd search.} Often a web search covers a broad spectrum and various facets, with all information unlikely to be found on a single web page, and thus often resulting in a lengthy process. Socializing with people with interests has the potential to speed up this process significantly. However, browsing and searching the Web is still mainly a single-user task, with no support for socializing in an ad-hoc manner.

With the advent of the Web 2.0, online fora or Q\&A systems, but also commenting sections on web pages, allow users to get in contact with each other and to share information. However, the communication is asynchronous, i.e., users typically have to wait quite a while to get a reply. Thus, the degree of socializing is still rather limited, which is particularly pronounced for questions that need or benefit from a quick answer. Furthermore, communities in such systems form in an explicit fashion: users have to join an online forum, or have to create and maintain a personal profile that reflects their interests and expertise. This rather static group formation does not well reflect the often ad-hoc information needs of a user.

In this paper, we propose the concept of presence on the Web as a means to allow user to socialize and exchange information in an ad-hoc manner. The intuition is that the web page a user is visiting is a good indicator for the user's interests. Thus, two users browsing the same page at the same time are likely to share common interests or looking for similar information. Our approach enables users to be aware of each other's presence and provides them with means to contact each together. To do this, we define the concept of a \textit{virtual location} -- as counterpart to a physical location.
\begin{myExample}
  Consider two users having encountered, independently, the same bug in a program. Using the error code and/or some keywords describing the error as input for a search engine is likely to bring both users on the same web page addressing that error. Now being aware of each other, the users can get in contact with each other, e.g., sharing their (successful or unsuccessful) attempts to find a solution.
\label{exp:web2web}
\end{myExample}

As a second major goal, we also aim to merge the concepts of presence in the physical and virtual space.  Many real-world entities feature a physical as well as a virtual location. For example, a hotel has a physical location specified by a geo coordinate and a virtual location, e.g., the hotel's website. With the advances in mobile technologies, people are able to be online almost everywhere. Sharing the presence of people among the virtual and physical space enables entirely new ways to interact, particularly in the context in-situ searches.
\begin{myExample}
  Imagine a user looking for a hotel room for the upcoming night. The websites of potential hotels provide the user with the usual information, e.g., rate per night, facilities, etc. Hotel recommendation sites provide additional information like the general quality of the rooms or the breakfast. Other information, however, e.g., if there is a noisy construction site close by, are usually not available, yet often important. If a user could contact current guest of a hotel, this could be solved.
\label{exp:web2geo}
\end{myExample}
Presence is defined as the part of the space within one's immediate vicinity. Besides \textit{space}, this definition involves the concepts of \textit{location} as one's current position, and \textit{distance} as a means to quantify the extent of one's vicinity. These concepts are well understood in the physical world, but much less so in the virtual one. We argue that the concept of a virtual presence goes beyond traditional understandings of presence. We then define the notions of location, distance, and presence on the Web within a formal framework. To demonstrate its feasibility, we implemented a proof-of-concept prototype. Privacy of presence and similar information tied to the identity of users and their behavior is a key concern. We are aware of this. However, before privacy can be addressed, a system needs to be modeled and understood, so that the implications on privacy can be assessed and its trade-off with the usefulness and usability of a system can be analyzed.

Paper Outline: Section~\ref{sec:related_work} reviews related work. Then Section~\ref{sec:physical_vs_virtual} outline the challenges towards the notion of a virtual presence. Section~\ref{sec:presence} presents our formal model based on the definitions of virtual location, virtual distance, and virtual presence; Section~\ref{sec:implementation} describes our proof-of-concept implementation. Section~\ref{sec:evaluation} evaluates the practical relevance and potential benefits of the presented approach. Finally, Section~\ref{sec:conclusions} concludes and outlines future work.

\section{Related Work}
\label{sec:related_work}

\textit{Browsing behavior of users.}  There are two principle approaches to investigate users' browsing behavior: \textit{(a) Server-side studies} such as~\cite{Xue10UserNavigation,Meiss09WhatsInASession} analyze server access logs. \cite{Xue10UserNavigation} shows that users often feature different behavior patterns, rather than a single one. 
The results in \cite{Meiss09WhatsInASession} confirm previous findings about long-tailed distributions in site traffic. \cite{Agichtein06ImprovingWebSearch,Wedig06LargeScaleAnalysis} analyze search engine interaction logs to gain insights into the query behavior. \cite{Agichtein06ImprovingWebSearch} investigated how users' search behavior can be used as feedback to improve the ranking of query results. 
\cite{Wedig06LargeScaleAnalysis} found that after a few hundred queries a user's topical interest distribution converges. \textit{(b)~Client-side studies} collect data on the client side using, e.g., browser plugins that log all user actions. \cite{Goel12WhoDoesWhat} focused on demographic factors, i.e., how age, sex, etc. affect users' browsing behavior. In \cite{Adar08LargeScaleAnalysis} the authors identify different types of revisitation behavior, providing recommendations towards web browser, search engines, and web design. \cite{Dubroy10AStudyOfTabbed,Zhang11MeasuringWebPage} investigated in detail tabbed browsing, i.e., the benefits of multiple tabs within a browser window. 

\textit{Similarity measures for web resources.}  Extending presence across similar virtual locations essentially translates to the identification of similar web pages.  \textit{(a) Content-based techniques} such as \cite{DaZhen07NearReplicas,Shivakumar98FindingNearReplicas} calculate so-called fingerprints derived from the content of pages. The similarity of two pages is derived from the similarity of the corresponding fingerprints. \cite{Broder97SyntacticClustering} chunks the page content into \textit{shingles}, i.e., contiguous subsequences of terms. The degree of overlap and/or containment between two sets of shingles quantifies the similarity between the pages. \textit{(b) Link-based techniques} consider two pages as similar if they are similarly embedded in the Web graph. A baseline algorithm is \textsc{SimRank}~\cite{Jeh02SimRank} where the similarity between two objects (e.g., web pages) is recursively defined as the average similarity between the objects linked with them. \textsc{SimRank} spurred 
the development of further extensions and refinements, such as \textsc{SimRank++}~\cite{Antonellis:2008:SQR:1453856.1453903}, 
or \textsc{MatchSim}~\cite{Lin09MatchSim}. \textit{(c) Structure-based techniques} deem two pages similar if they ``look similar''. \cite{Li05WebDataExtraction} directly compares the DOM tree of pages; \cite{Bohunsky10VisualStructureBased} first reduces the DOM tree to its visually most relevant components, and then compares them. \cite{Fu06DetectingFishing} converts web pages into images to perform an image analysis to detect similarities. \textit{(d) Combined measures} aim to combine the advantages of different analysis approaches, e.g., combining content- and link-based techniques. \textsc{WordRank}~\cite{Kritikopoulos07WordRank} extends \textsc{PageRank}~\cite{Page98PageRank}, introducing weights to put biases on the links to pages with similar textual content. \cite{Chakrabarti98Automatic} extends the \textsc{HITS}~\cite{Kleinberg99HITS} following a similar approach as \textsc{WordRank}. 

\textit{Collaborative browsing and searching.} Browsing and searching the Web is still primarily an isolated task. \cite{Morris08Survey} conducted a survey showing that collaborative browsing is crucial for many users, but currently requires users to revert to out-of-band channels such as phone, e-mail or instant messaging. \textsc{TeamSearch}~\cite{Morris06TeamSearch} and \textsc{CoSearch}~\cite{Amershi08CoSearch} are systems providing mechanisms for co-located collaboration, i.e., where several users gather around one computer. \textsc{SearchTogether}~\cite{Morris07SearchTogether} and \textsc{CoScripter}~\cite{Leshed08CoScripter} extend this idea to collaborative browsing between users working with their own computers. \textsc{COBS} (\textit{CO}llaborative \textit{B}rowsing and \textit{S}earching)~\cite{vdw11COBS} proposes a browser extension providing a proof-of-concept implementation that allows users visiting the same site to communicate with each other.

In summary, existing works do not consider the Web as a space in which user can not only navigate, but also ``meet'' at locations which potentially reflect their interests and likings, particularly in an ad-hoc manner. To improve the browsing experience of users by social aspects and enable novel ways to interact with and discover (implicit and explicit) information, we introduce the concepts of presence and space into the Web. To enable this, we provide a formal framework that adopts the well-defined notions of location, distance, and presence from the physical world and brings them into the virtual space of the Web in a meaningful way, so that the physical and the virtual worlds are presented to the user through uniform, easy to comprehend abstractions.

\section{Physical vs. Virtual Space}
\label{sec:physical_vs_virtual}
Despite sharing similar notions, physical and virtual presence feature fundamental differences affecting the design of presence awareness mechanisms on the Web. In the following, we use the term \textit{walker} to denote a user's identity in the physical
space and \textit{surfer} to denote it in the virtual space.

\textit{Structure of the virtual space.}  While the physical space is continuous, for its representation in a data repository discrete models are used~\cite{Gueting05Moving}. Regarding web page a a location, it is not meaningful to assign a user's location to a point between two pages. Thus, the virtual space is discrete, typically simplifying the definition of notions for a virtual presence awareness.

\textit{Distance and locality.}  In the physical space, the distance between two locations is well-defined, e.g., using the Euclidean distance. Between two web pages, in general, such distance measures are missing. While one can define the distance between two pages $u_1$ and $u_2$, e.g., as the minimum number of hyperlinks needed to be followed to get from $u_1$ to $u_2$, this does not necessarily constitute a meaningful distance definition in an application context.

\textit{Moving between locations.} 
Natural limitations regarding the the time regarding to move physical location as well as regarding possible directions often allows, to some extent, to anticipate a person's or object's location in the near future. On the Web, however, a surfer can navigate at any point in time to any page. Thus, to reliably predict the next page a user will navigate to is, in general, not possible.

\textit{Identity protection.} One's presence in the physical space is typically known to everybody in one's vicinity. Online, however, mechanisms to shield/hide one's personal data are omnipresent. In social networks, for example, users create explicit connections to others and organize them into groups. Applied to presence awareness, such mechanisms enable users to reveal their presence at a location only to selected users.

\textit{Symmetry.}  In the physical space, in general, if a person $A$ is aware of a person $B$, so is $B$ of $A$. Mechanisms supporting to shield one's
presence from others break this symmetry. Due to privacy concerns, surfers have a strong incentive to hide, potentially resulting in a majority of hiding
surfers. Obviosuly there is a trade-off between privacy protection mechanisms and incentives for users to actively contribute.

\textit{Multiple locations.}  At any time, a walker in the physical space always occupies one unique location. On the Web, a surfer can visit multiple web pages by using multiple browser windows or tabs. Thus, a surfer may have multiple presences and is aware of distinct groups of other surfers or walkers. Any techniques towards group collaboration, e.g., chat messaging, have to distinguish between such groups.

\section{Presence on the Web}
\label{sec:presence}
Our approach to adopt the notion of presence to the Web involves (a) the definition of required concepts, such as virtual location and virtual distance, and (b) different techniques that decide, if and when two users are aware of each other.

\subsection{Virtual Location and Distance}
Simply speaking, space describes the possibilities where a person ``can be''. Given these notions, we can define the virtual space as the set of websites a user can visit.
\\
\\
\textbf{Virtual coordinates and locations.}  In geographic terms, the most fine-grained way to specify a walker's current position is by geo coordinates, e.g., longitude and latitude. Mapping this concept to the virtual space, the current position of a user is the web page the user is visiting. Thus, within our framework, each resource on the Web is uniquely identified by a URI.
\\
\\
\begin{myDefinition}{Virtual coordinate}
A virtual coordinate $vc$ is the URI of a web page.
\end{myDefinition}

In many application contexts, not the distinct page but the category or topic or similar concepts of a page are of interest to describe a surfer's location. We therefore extend the definition of a virtual location beyond a single virtual coordinate.
\\
\\
\begin{myDefinition}{Virtual location}
A virtual location $vl$ is a distinct, non-empty, and finite set of virtual coordinates $\mathcal{V} = \{vc_1, vc_2, ..., vc_n\}$, with $\mathcal{V}_1 \cap \mathcal{V}_2 = \emptyset$.
\end{myDefinition}
What virtual coordinates constitute a virtual location, e.g., representing a topic, is application-specific.
\\
\\
\textbf{Virtual distance.}  Vicinity as part of the definition of presence requires a notion of distance. With the goal to bring together surfers with similar interests or likings, we define the distance between pages by means of their similarity. For this, we utilize existing efforts to quantify the similarity between web pages (see Section~\ref{sec:related_work}). To combine similarity measures, we assume each measure $sim_k(vl_i, vl_j)$ between two locations $vl_i$, $vl_j$ to be normalized, i.e., $sim_k(vl_i, vl_j)\in [0, 1]$.
\\
\\
\begin{myDefinition}{Virtual distance.}
The virtual distance between two virtual locations $vl_i$ and $vl_j$ is the weighted average of the results of a set of existing, normalized similarity measures:

\smallskip
\hfil
$d_{web}(vl_i, vl_j) = \sum_k { w_k \cdot sim_k(vl_i, vl_j)\ , \quad
  \mathrm{with}\ \sum_k{w_k}=1  }$
\hfil
\smallskip

\noindent Note that $d_{web}(vl_i, vl_j)\in [0,1]$.
\end{myDefinition}
Depending on the applied measures and the values of $w_k$, two virtual locations are close if they cover similar topics, feature similar tags, are similarly embedded in the Web graph, etc. 

\subsection{Virtual Presence}
The traditional definition of presence involves that people are aware of each other if they are at same location at the same time. The nature of the Web in turn allows us to go beyond the notion of one's \textit{immediate} vicinity. Intuitively, the overall presence of a user at a location depends on two aspects. Firstly, it depends on the overall time the user spends at that location derived from the number of visits and the length of the individual visits. Secondly, recent visits typically have more impact on a user's presence than visits further in the past.

Let $\mathcal{U}$ denote the set of surfers, and $u\in \mathcal{U}$ be a individual surfer. $\mathcal{L}$ is the
set of all virtual locations, and $vl\in \mathcal{L}$ is a single location. $V(u,vl)$ denotes the set of visits of a user $u$ at $vl$, $v\in V(u,vl)$, with
$v=[v_s, v_e]$, denotes the interval of a individual visit, from time $v_s$ till $v_e$. With this, the time $\tau_{u,vl}$ user $u$ was at location $vl$ can be calculated as 
\begin{equation}
\tau_{u,vl}\!=\!\sum\limits_{v=[v_s, v_e]\in V(u,vl)}^{} [v_e\!-\!v_s ]\ = \sum\limits_{v=[v_s, v_e]\in V(u,vl)}^{}\!\left( \int\limits_{v_s}^{v_e} \mathbf{1}\ dt \right)
\label{formula:dynamic01}
\end{equation}
To emphasize recent visits we introduce the notion of a \textit{decay function} to the impact of past visits on the overall visit time.
\\
\\
\begin{myDefinition}{Decay function}\\
\noindent \begin{minipage}{0.45\linewidth}
    \noindent Let $\delta^\prime(t)$ be a strictly monotonically decreasing function. We now define a decay function $\delta(t)$ as:
  \end{minipage}
  \hspace{0.06\linewidth}
  \begin{minipage}[]{0.48\linewidth}
    \begin{equation}
\delta(t)=\begin{cases}
  1 &,\ \delta^\prime(t) > 1\\
  \delta^\prime(t) &,\ 0 \leq \delta^\prime(t) \leq 1\\
  0 &,\ \delta^\prime(t) < 0
\end{cases}\\
\end{equation}
\end{minipage}\\
\noindent The distinction of cases ensures that $\delta(t)\in [0,1]$. Note
that a meaningful decay function $\delta(t)$ requires that $\delta^{\prime}(t)
\in [0,1]$.
\end{myDefinition}

\noindent \begin{minipage}{0.45\linewidth}
  \noindent To integrate the decay function $\delta(t)$ into the visiting time $\tau_{u,vl,\delta(t)}$
  of user $u$ at location $vl$ we extend Formula~\ref{formula:dynamic01}
  with $\delta(t)$ as follows:
  \end{minipage}
  \hspace{0.03\linewidth}
  \begin{minipage}[]{0.51\linewidth}
    \begin{equation}
\tau_{u,vl,\delta(t)} = \sum\limits_{v=[v_s, v_e]\in V(u,vl)}^{} \left( \int\limits_{v_s}^{v_e} \delta(t)\ dt \right)
\label{formula:dynamic02}
\end{equation}
\end{minipage}
\\
\\
We can now define the presence of a user $u$ at location $vl$ by normalizing $\tau_{u,vl,\delta(t)}$ with respect to the decay function
$\delta(t)$:

\begin{myDefinition}{Presence}
The presence $p(u,vl,\delta(t))$ of a user $u$ at a location $vl$, given a decay function $\delta(t)$, is defined as:
\begin{equation}
p(u,vl,\delta(t))\!=\frac{\tau_{u,vl,\delta(t)}} {\int\limits_{t_{-\infty}}^{t_{now}} \delta(t)\ dt}\!=\!\frac{\sum\limits_{v=[v_s, v_e]\in V(u,vl)}^{} \left( \int\limits_{v_s}^{v_e} \delta(t)\ dt \right)}{\int\limits_{t_{-\infty}}^{t_{now}} \delta(t)\ dt}
\end{equation}
Note that $p(u,vl,\delta(t))\in [0,1]$ and that $p(u,vl) = 1$ if $u$ was at $vl$ during the whole considered time span. Figure~\ref{fig:example_presence} shows an example for the evaluation of $p(u,vl,\delta(t))$.
\label{def:presence}
\end{myDefinition}

\noindent \begin{minipage}[]{0.5\linewidth}
  \noindent \begin{myExample} Figure~\ref{fig:example_presence} illustrates the
    evaluation of the weighted visiting time of a user $u$ at a virtual
    location $vl$: $u$ has visited $vl$ three times in the last hour (marked
    areas). We use an exponential decay function $\delta(t) = e^{\lambda t}$,
    with $\lambda = -0.05$. The marked areas represent the weighted visiting
    time $\tau_{u,vl,\delta(t)}$. $p(u,vl,\delta(t))=0.328$ is the ratio between the marked
    areas and the whole area below the decay function $\delta(t)$. 
    \label{exp:example_presence}
    \end{myExample}
  \end{minipage}
  \hspace{0.2cm}
  \begin{minipage}[]{0.48\linewidth}
    \centering
    \includegraphics[width=1\textwidth]{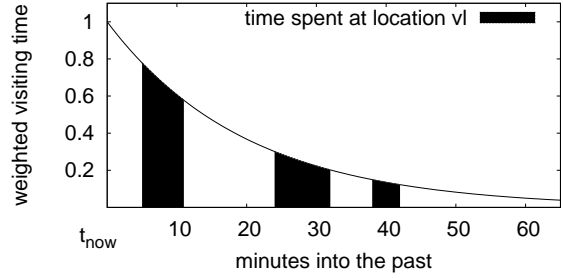} 
    \captionof{figure}{Example for virtual presence}
    \label{fig:example_presence}
  \end{minipage}

We now extend the notion of presence by the time spent at similar locations with their impact on the overall presence value depending on their similarity/closeness to $vl$. At first, we assume that each user is present only at one location at a time.

\begin{equation}
\hat \tau_{u,vl,\delta(t)}\!=\!\sum\limits_{l_i \in L}^{} \left[ \sum\limits_{v\!=\![v_s, v_e] \in V(u,l_i)}^{} \left( \int\limits_{v_s}^{v_e} d_{web}(vl,vl_i)\delta(t)\ dt \right) \right]
\label{formula:dynamic03}
\end{equation}
Analogously, we can now define the notion of the \textit{cumulative presence} of users at specific locations by normalizing $\hat \tau_{u,vl,\delta(t)}$ with respect to $\delta(t)$.
\\
\\
\begin{myDefinition}{Single-location cumulative presence}\\
\noindent \begin{minipage}{0.45\linewidth}
    \noindent The cumulative presence $\hat p(u,vl,\delta(t))$ of a user $u$ at a virtual location $vl$, with a decay function $\delta(t)$, is defined as:
  \end{minipage}
  \hspace{0.06\linewidth}
  \begin{minipage}[]{0.48\linewidth}
\begin{equation}
\hat p(u,vl,\delta(t)) = \frac{\hat \tau_{u,vl,\delta(t)}}{\int\limits_{t_{-\infty}}^{t_{now}} \delta(t)\ dt}
\end{equation}
\end{minipage}\\
\noindent Note that also $\hat p(u,vl,\delta(t))\in [0,1]$ since $u$ is only present at one location at a time. 
\end{myDefinition}

\noindent  \begin{minipage}[]{0.5\linewidth}
    \noindent \begin{myExample}
    We extend Example~\ref{exp:example_presence}. Here we not only consider
    user $u$'s visits at location $vl$ but also $u$'s visits at a related
    location $vl^\prime$. The similarity between both locations is
    $d_{web}(vl,vl^\prime) = 0.75$. The effect of the time spent at $vl^\prime$
    is represented by the checkered areas in Figure~\ref{fig:example_presence_cummulative}. Note that the height of the checkered areas are $0.75$-times the height of the upper bound defined by $\delta(t)$, reflecting $d_{web}(vl,vl^\prime)=0.75$. Here, $ \hat p(u,vl,\delta(t)) = 0.457$.
    \end{myExample}
  \end{minipage}
  \hspace{0.2cm}
  \begin{minipage}[]{0.48\linewidth}
    \centering
    \includegraphics[width=1\textwidth]{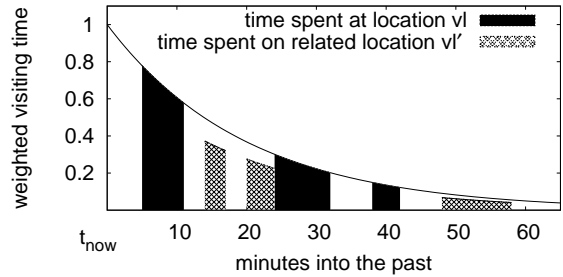}
    \captionof{figure}{Example for single-location cumulative presence}
    \label{fig:example_presence_cummulative}
  \end{minipage}

A surfer might be present at more than one location at a time, for example, when surfing the Web with two open browser windows. This case includes that a surfer can be at the same time at two locations that are similar. Here, simply summing up the individual visit time would distort the results. To ensure comparable cumulative presence values for a specific user at a specific location, an adequate definition of cumulative presence is required. We therefore rewrite Definition~\ref{formula:dynamic03} to incorporate overlapping presences at similar locations.

\begin{equation}
\hat{ \hat \tau}_{u,vl,\delta(t)}\!=\!\sum\limits_{v\!=\![v_s, v_e] \in
  V(u,vl_i)}^{} \left( \int\limits_{v_s}^{v_e} max\left(\bigcup f(t,u,vl,vl_i)
  \right) \delta(t)\ dt \right)\nonumber
\end{equation}
\vspace{-0.4cm}
\begin{equation}
f(t,u,vl,vl_i) = 
\begin{cases}
  d_{web}(vl,vl_i) &,\ t\in V(u,vl_i) \\
  0 &,\ \text{otherwise}
\end{cases} 
\label{formula:dynamic04}
\end{equation}
$t\in V(u,l_i)$ is true if there is a visit interval $[v_s,v_e] \in
V(u,vl_i)$ with $t\in [v_s,v_e]$. To finally define the notion of a
multiple-location cumulative presence we, again, normalize $\hat{ \hat
  \tau}_{u,vl,\delta(t)}$ with respect to the decay function $\delta(t)$.

\begin{myDefinition}{Multiple-locations cumulative presence}\\
\noindent \begin{minipage}{0.45\linewidth}
    \noindent The cumulative presence $\hat{ \hat \tau}(u,vl,\delta(t))$ of a user $u$ on a  virtual location $vl$, with a decay function $\delta(t)$, is defined as:
  \end{minipage}
  \hspace{0.06\linewidth}
  \begin{minipage}[]{0.48\linewidth}
   \begin{equation}
 \hat{ \hat p}(u,vl,\delta(t)) = \frac{\hat{ \hat \tau}_{u,vl,\delta(t)}}{\int\limits_{t_{-\infty}}^{t_{now}} \delta(t)\ dt}
\end{equation}
\end{minipage}
\end{myDefinition}

\noindent \begin{minipage}[]{0.5\linewidth}
    \noindent \begin{myExample}
    Figure~\ref{fig:example_presence_cummulative_multi} illustrates
    the case where a user has visited two related virtual locations
    $vl$ ad $vl^\prime$ at the same time. Again, the checkered areas
    represent the time spent on a related location with
    $d_{web}(vl,vl^\prime) = 0.75$. Compared to
    Figure~\ref{fig:example_presence_cummulative}, however, here the
    visits at both locations overlap. Formula~\ref{formula:dynamic04}
    ensures that the parallel visits of related locations do not
    over-emphasize a user's presence at a location. Here, $\hat{ \hat
        p}(u,vl,\delta(t)) = 0.522$.
    \end{myExample}
  \end{minipage}
  \hspace{0.2cm}
  \begin{minipage}[]{0.48\linewidth}
    \centering
    \includegraphics[width=1\textwidth]{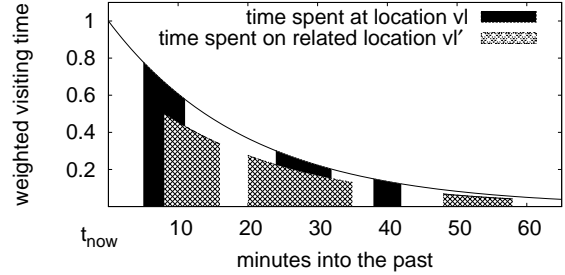}
    \captionof{figure}{Example for multiple-locations cumulative presence}
    \label{fig:example_presence_cummulative_multi}
  \end{minipage}

\subsection{Presence Awareness}
\begin{SCfigure}
  \centering
  \includegraphics[width=0.75\textwidth]{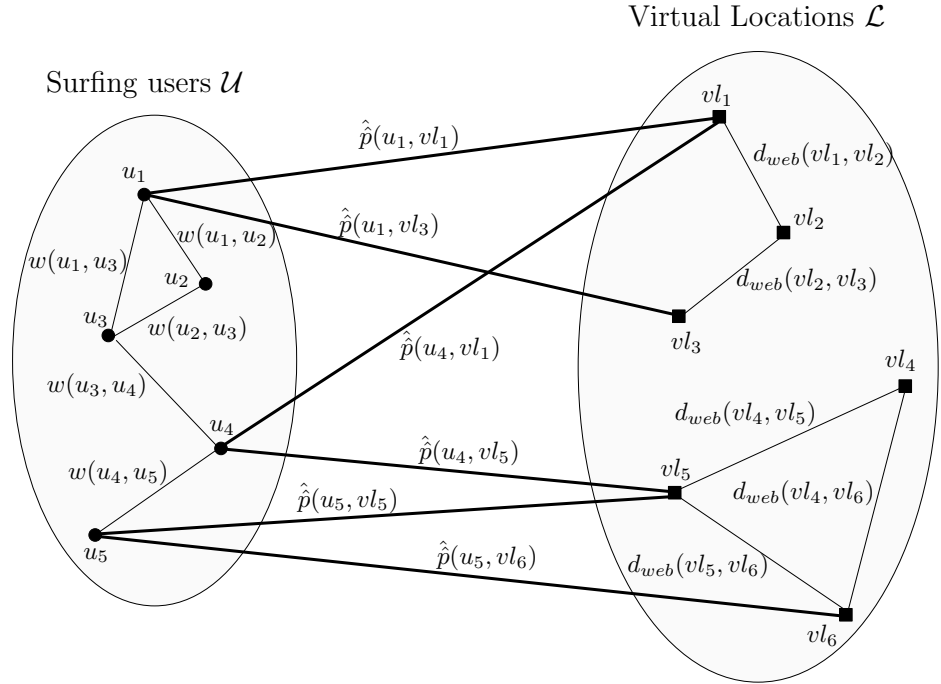}
\caption{Graph representation of the underlying data model showing users, virtual locations, and the relationships between them.}
\label{fig:bipartite_example}
\end{SCfigure}

Most intuitively, two surfers ``meet'' if they are ot the same virtual location at the same time. Our definition of presence also allows for more sophisticated schemes to decide if and when a surfer is aware of others. A straightforward application is to rank or filter surfers that are currently at the same location according to their presence values. The more interesting application of presence values is their use to extend presence awareness to users that are currently not at the same virtual locations. Figure~\ref{fig:bipartite_example} depicts the general underlying model as a graph $G(V,E)$. The set of vertices $V = \mathcal{U} \cup \mathcal{L}$ (with $\mathcal{U} \cap \mathcal{L} = \emptyset$) is the union of the set of surfers $\mathcal{U}$ and the set of virtual locations $\mathcal{L}$. $E = E_{\mathcal{U}} \cup _{\mathcal{L}} \cup E_{\mathcal{UL}}$, where $E_{\mathcal{U}}$, $E_{\mathcal{L}}$, $E_{\mathcal{UL}}$ are pairwise disjoint. $E_{\mathcal{U}} = \{e\!=\!(u_i,u_j)\in E |\ u_i, u_j \in \mathcal{
U}\}$ represents relationships that may exist between users. These relationships may form explicitly, like contacts in social networks or implicitly by means of, e.g., trust or reputation mechanisms. To simplify the presentation we assume symmetric relationships, thus resulting in undirected edges between users. Edges between users may feature a weight or label to reflect the strength of the tie between the users, e.g., trust values. A weighted edge in $E_{\mathcal{L}} = \{e\!=\!(vl_i,vl_j)\in E |\
vl_i, vl_j \in \mathcal{L}\}$ represents virtual distance between two locations. Finally, $E_{\mathcal{UL}} = \{e\!=\!(u,vl)\in E |\ u\in \mathcal{U}\ \wedge\ vl \in \mathcal{L}\}$ connects users with locations based on the available presence information. The weight of the edge between a user $u$ and a location $vl$ directly derives from the cumulative presence value $\hat {\hat p}(u,vl)$.

A plethora of existing work in context of, among others, graph analysis / (social) network analysis, recommender systems, reputations systems, trust management can be applied to identify a meaningful set of surfers a user is aware of. From a practical point of view, the presence-specific information, $E_{\mathcal{UL}}$ and respective edge weights, can be very dynamic depending on the user's actions. The frequent periodic application of algorithms on the
whole or large portions of the presence graph is no practicable at large scale. Thus, we will investigate feasible approximations in our on-going work.

\section{Implementation}
\label{sec:implementation}
\begin{SCfigure*}
  \centering
  \includegraphics[width=0.75\textwidth]{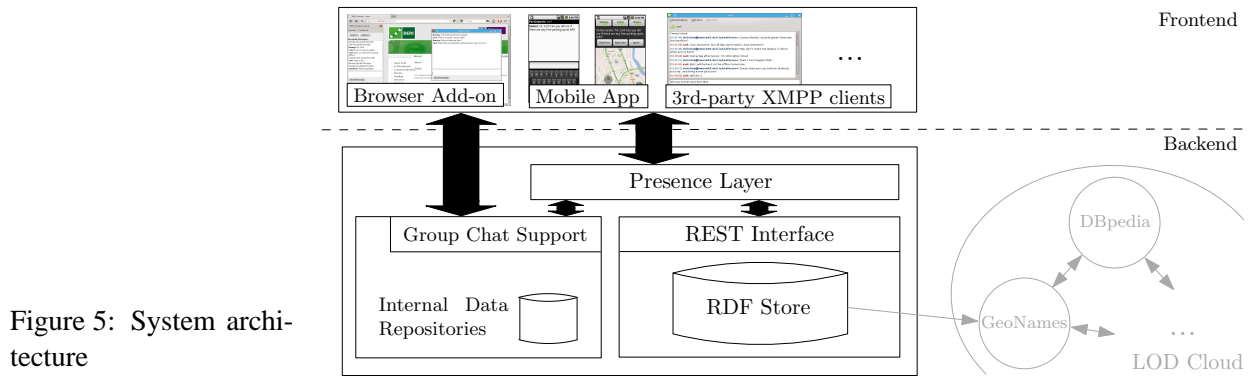}
  \caption{System architecture}
  \label{fig:architecture}
\end{SCfigure*}

In this section we present our current proof-of-concept implementation. As a first large-scale, real-world deployment, our implementation was part of the official mobile app\footnote{http://www.nuigalway.ie/volvo-ocean-race-2012/deriapp/} for the Volvo Ocean Race 2011-2012.
Figure~\ref{fig:architecture} shows the overall system architecture, described in the following.

\subsection{Backend Architecture}
The backend consists of three components: an RDF repository to store and associate physical and virtual locations, a RESTful interface to access
the repository, and an XMPP server for the exchange of presence information and messages.

\textbf{RDF Repository and RESTful Interface.} The repository maintains the mapping between the physical and virtual locations. We represent physical locations using 2-dim\-en\-sio\-nal geo coordinates, i.e., latitude and longitude.  For the time being, we focus on ``single point'' locations like hotels, hospitals, pubs, shops, etc. We also store locations with a large spatial extent, like parks or golf courses, using single geo coordinates. 
The data collection process is described Section~\ref{sec:evaluation-s2w}. This data set can be accessed via a RESTful interface.

\textbf{XMPP Server.}  We provide presence awareness based on the open-standard Extensible Messaging and Presence Protocol (XMPP: http://xmpp.org). It supports user-to-user chats and group chats. Our backend features an XMPP server as the core component which enables any third-party solutions (e.g., instant messaging clients) with XMPP support to connect. Within our architecture, we assign each location, both physical and virtual, to a group chat. The intuition is that users at the same location are in the same group chat and are therefore aware of each other. Since walkers and surfers can have different physical and
virtual locations we distinguish between (a) \textit{geo group chats} representing the physical representation of a location and (b) \textit{web group chats} representing the location's virtual representation. For example, a guest in a hotel is not necessarily at the corresponding virtual location, i.e., the website of the hotel. Besides presence information, we enable users to exchange messages.We currently support group chats and the user-to-user communication between users.

\textbf{Presence Layer.} The presence layer connects the XMPP server with the data repository and handles the mapping of virtual locations stored in the repository onto group chats residing within the XMPP server. Once the group chat is determined and the user has entered it, any further communication is done directly via the XMPP server.

\subsection{Frontend}
\label{subsec:frontend}
\begin{SCfigure*}
  \centering
  \includegraphics[width=0.7\textwidth]{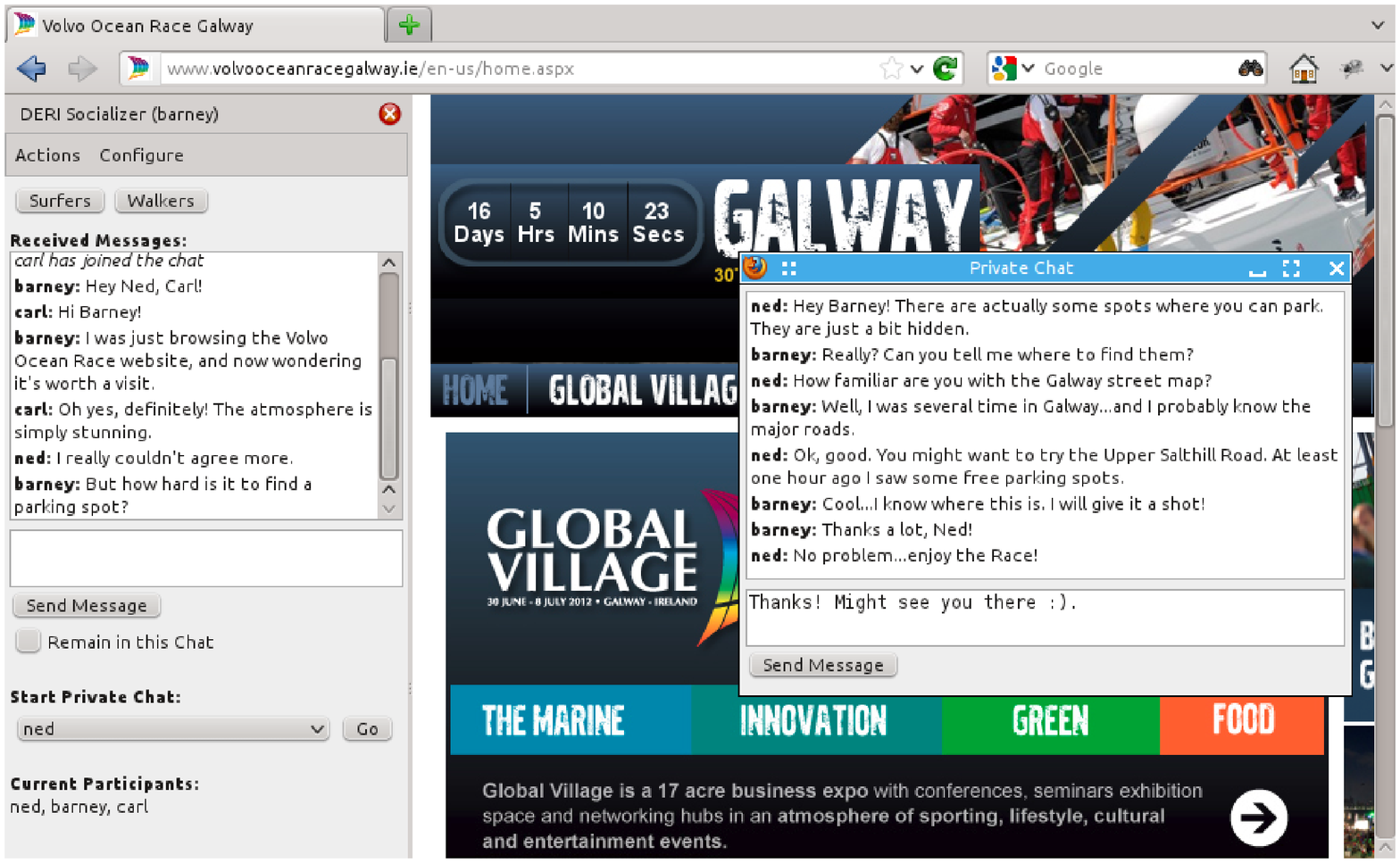}
  \caption{Example screenshot of the browser add-on showing (a) the group chat of virtual location of the Volvo Ocean Race, and (b) a private chat between two users.}
  \label{fig:addon}
\end{SCfigure*}

\begin{SCfigure*}
  \centering
  \includegraphics[width=0.20\textwidth]{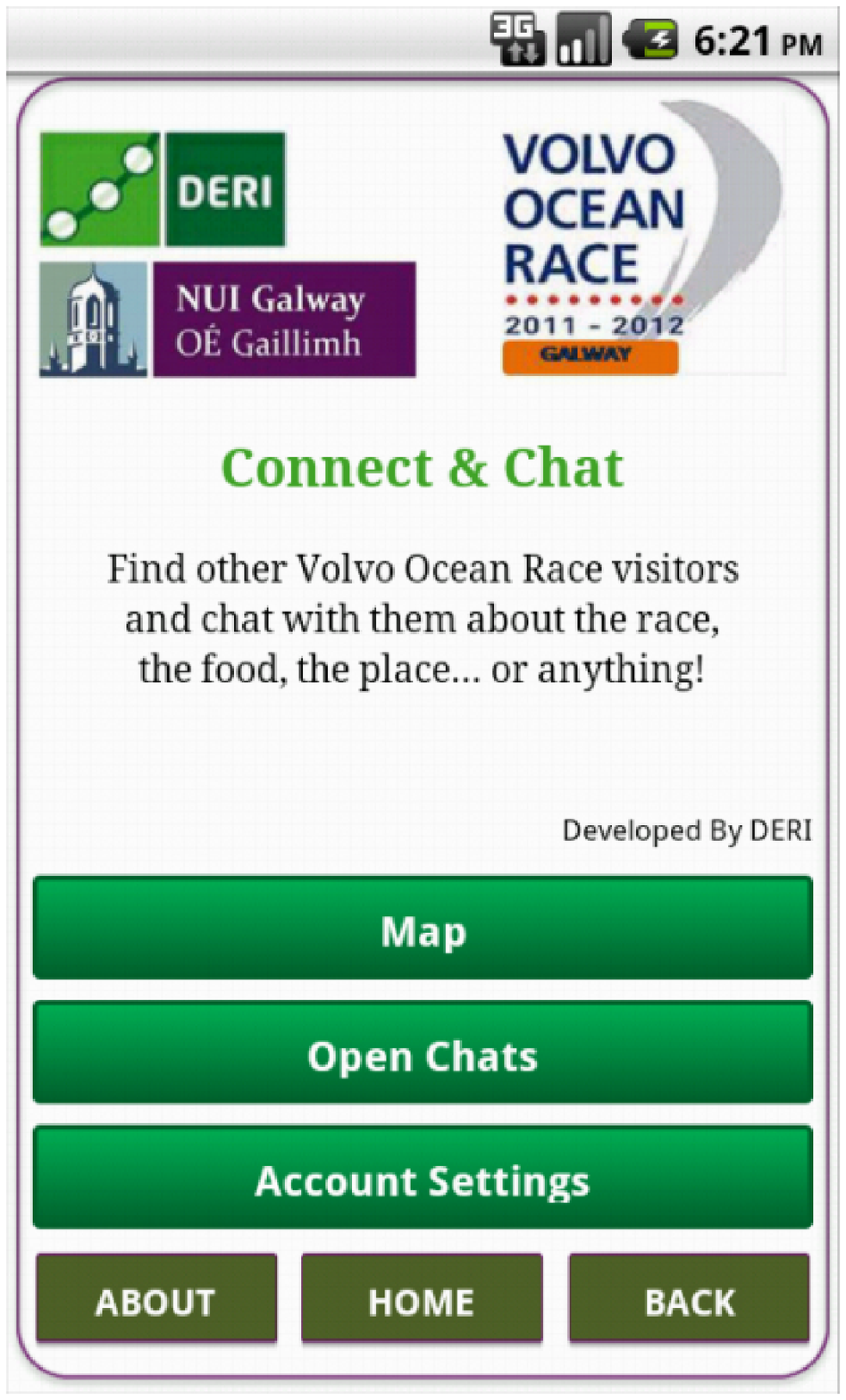}\quad
  \includegraphics[width=0.20\textwidth]{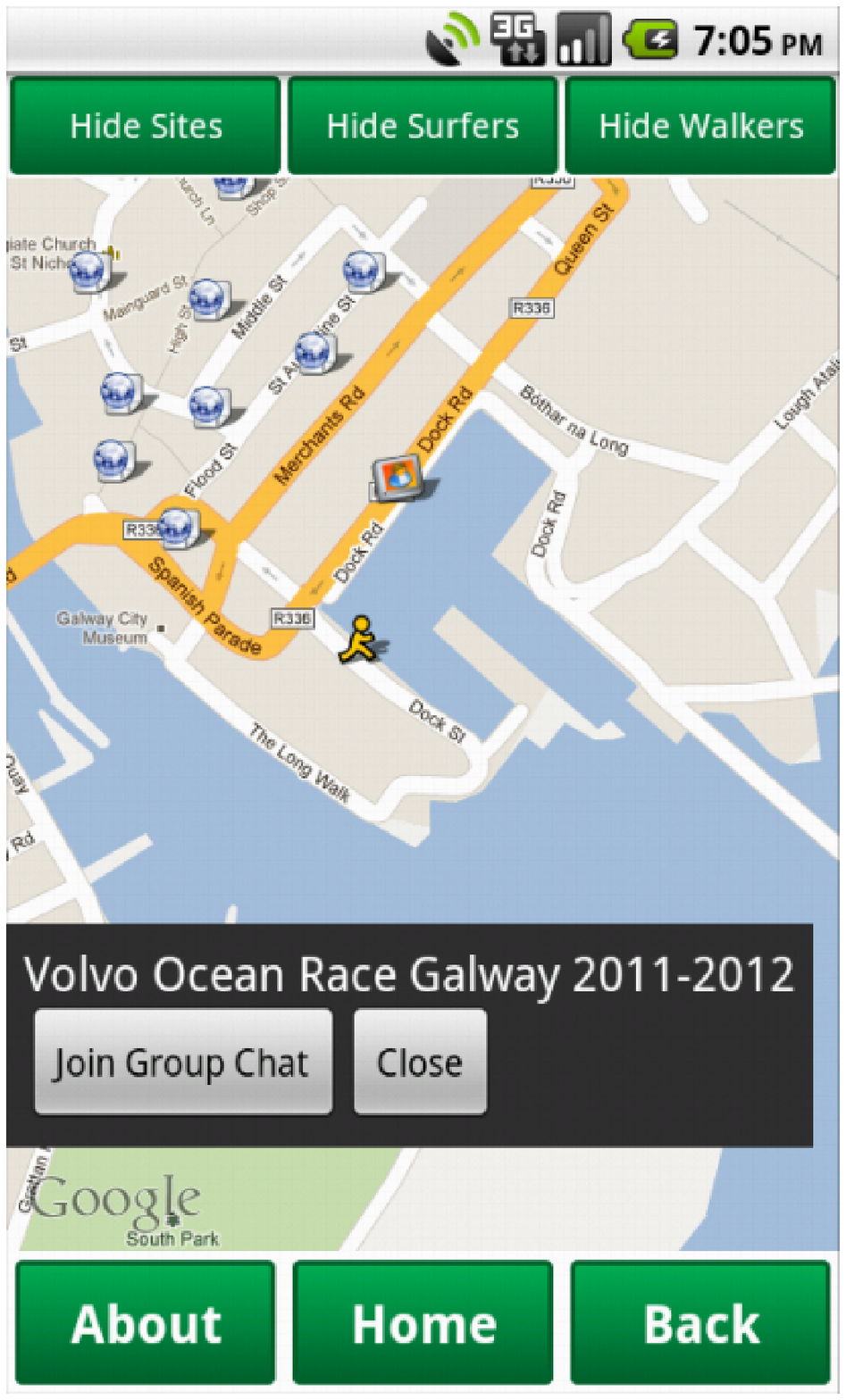}\quad
  \includegraphics[width=0.20\textwidth]{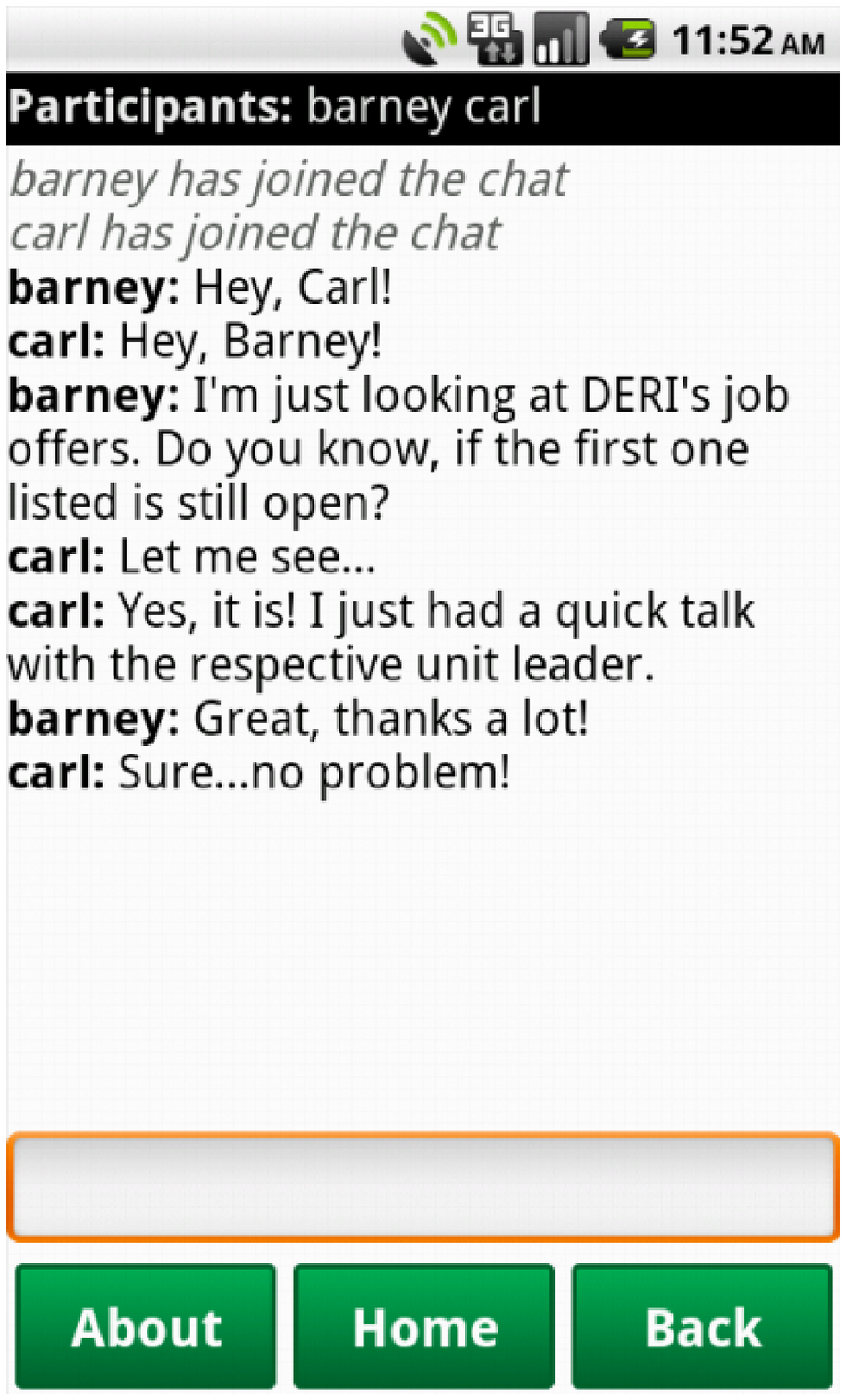}
  \caption{Mobile phone application. \textit{Left:} start screen for the Volvo Ocean Race. \textit{Middle:} map view showing virtual locations, surfers, walkers. \textit{Right:} chat view for private and group chats.}
  \label{fig:android}
\end{SCfigure*}

We now describe the frontend components of our current implementation.\footnote{More screenshots and use case descriptions available at http://vmusm02.deri.ie/presence} Since there are basic differences between moving in the physical and the virtual space, the interfaces for walkers and surfers must reflect this.
\\
\\
\textbf{Web browser add-on.} 
We aim for a seamless integration of our presence mechanisms into the normal browsing experience of users. We therefore implemented a browser add-on featuring a sidebar to display presence information in an unobtrusive manner (see Figures~\ref{fig:addon}). The add-on maintains an XMPP connection with references to the web group chat of the currently visited page, and to the geo group chat of the corresponding physical location (if available / applicable). The window for the web group chat allows users to communicate in a chat-like fashion: Users can send message to the group chat which can be read by all current participants in the room. Newcomers can read the most recent messages when entering the chat. 
\\
\\
\textbf{Mobile application.} We implemented a mobile application -- see Figures~\ref{fig:android} -- with two main features. The first one is a map based on the \textsc{Google Maps} API displaying all available virtual locations, surfers and walkers. Clicking on a location, surfer or walker, shows basic information about them. This dialog window also includes a button that allows the walker to enter the corresponding web group chat. The second feature is a basic chat client for private and group chats. Every time the physical location changes, the application sends a REST request to the backend with the new geo coordinates. The response is the closest virtual location in a radius of less than, e.g.,
100m. If such a location exists, the application enters the corresponding geo group chat (thus making walkers visible to surfers). 

\section{Evaluation}
\label{sec:evaluation}
While the notion of virtual presence is meaningful on a conceptual level, it still needs to be shown that this approach is also meaningful in practical terms. This essentially relates to the question, whether two or more users are indeed likely meet in well-defined locations both virtual and physical.

\subsection{Surfer-to-Surfer}
To quantify the probability that two or more surfers meet on the same page, we analyzed the page view statistics of Wikimedia as real-world dataset projects\footnote{http://dumps.wikimedia.org/other/pagecounts-raw/}. These statistics record how often a page has been requested within the timeframe of one hour. Each record contains the number of requests within a clock hour (0-1am, 1-2am, ...) We focused on articles in the English Wikipedia, and analyzed three days: Sept. 13th, 2008, 2010 and 2012. 
Figure~\ref{fig:visitor-distribution} shows the results. As expected, the distributions of visitors on Wikipedia articles follow a power law, i.e., only a small number of articles is read by a large number of users. For a more illustrative representation of the results, Figure~\ref{fig:meeting_probabilities} shows the average probability that $x$, with $x\in[2, 30]$, or more users requested the same page within an hour. Two major points are worth mentioning:

\begin{figure*}
\parbox{.48\linewidth}{
  \centering
  \includegraphics[width=0.48\textwidth]{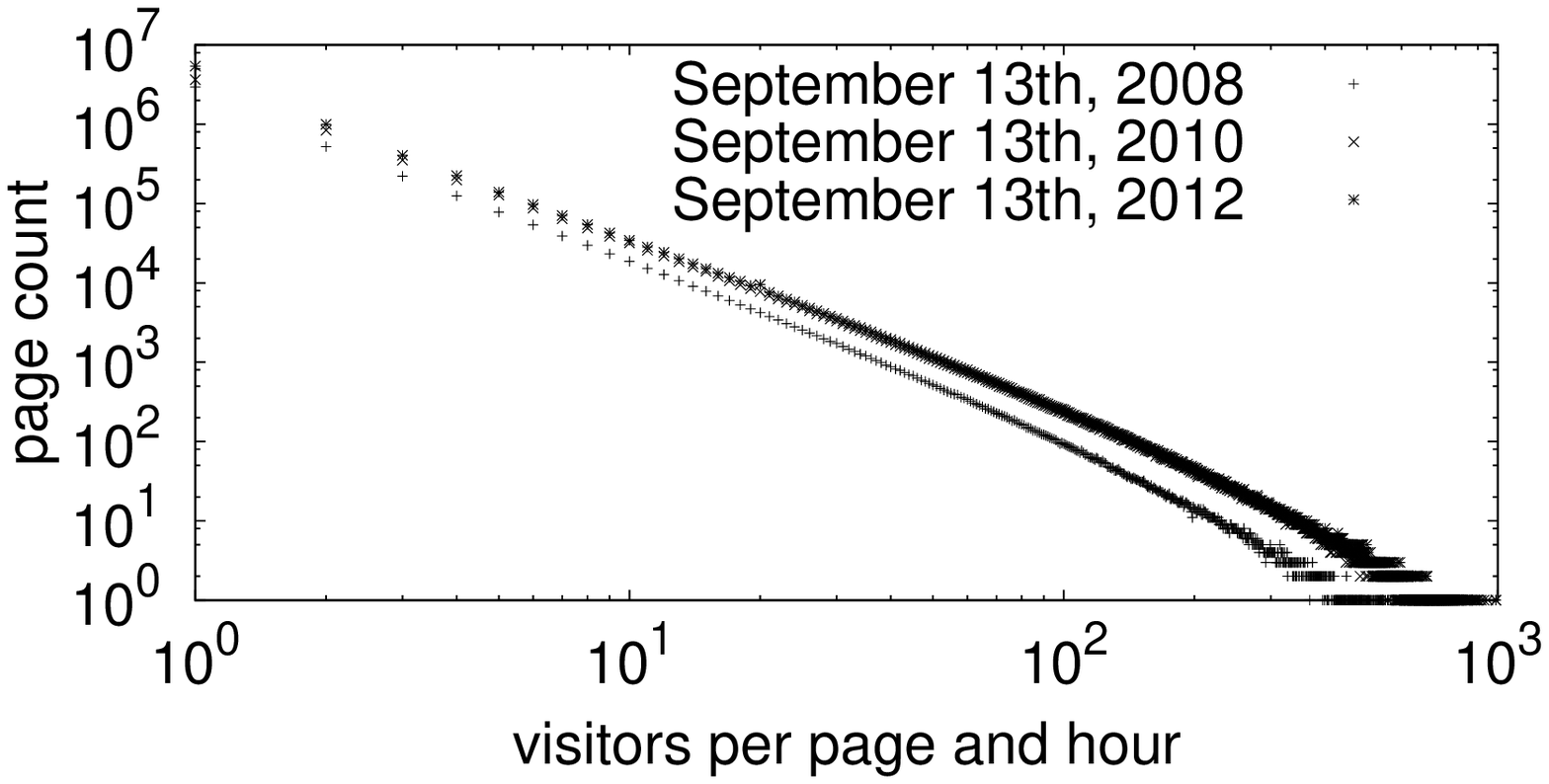}
  \caption{Wikipedia: visitor distribution.}
  \label{fig:visitor-distribution}
}
\hfill
\parbox{.48\linewidth}{
  \centering
  \includegraphics[width=0.48\textwidth]{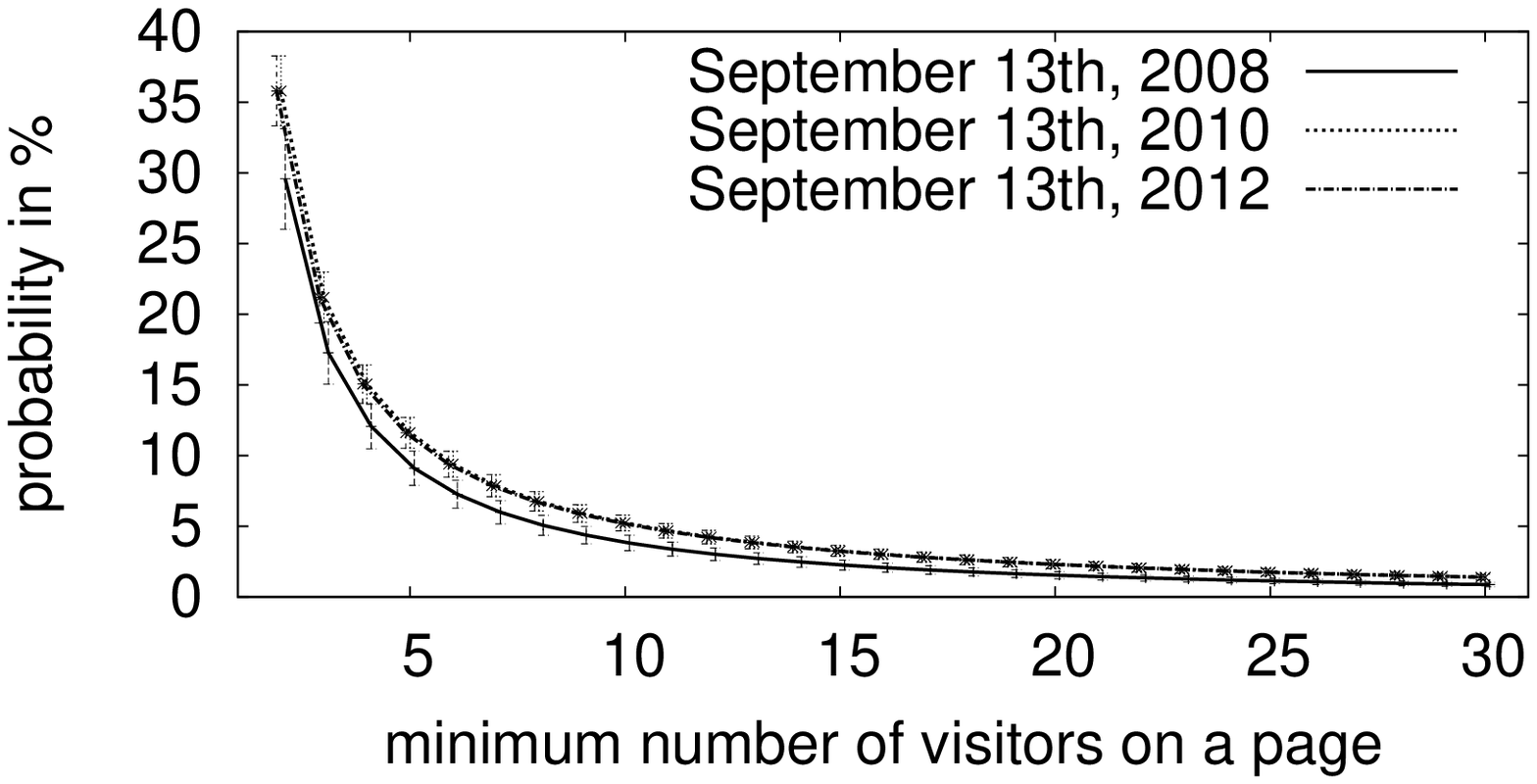}
 \caption{Wikipedia: meeting probabilities.}
 \label{fig:meeting_probabilities}
}
\end{figure*}

(1) The situation where two or more users visit the same page is far from being unlikely. This backs up our initial hypothesis that a presence mechanism potentially yields a real benefit for users. Thus, the Web does indeed represent a space allowing users to form implicit networks with other users by meeting and interacting with them on web pages.

(2) The probability that many users meet on a web page is not negligibly low. While this is a worthwhile situation, a
fruitful communication or collaboration with others is limited to a manageable number of parallel interactions. Suitable solutions for presence awareness and
ad-hoc communication must (a) provide mechanisms that allow users to rank, filter or group users, and (b) support different modes of communication, e.g.,
user-to-user or to a specified subset of users. 

(3) The probability for people to request the same article has slightly increased over time. The explanation for this is that the growth on the number of requests exceeds the growth of the number of requested pages. Simply speaking, significantly more users request only slightly more pages over time.

\subsection{Surfer-to-Walker}
\label{sec:evaluation-s2w}
In the following series of experiments we investigated if the folding the physical and virtual space provides a sufficient overlap to be of practical relevance. We collected two types of data: static data in terms of virtual locations with the geo coordinates of their physical counterparts, and dynamic data in terms of recorded GPS tracks. Table~\ref{tab:evaluation-parameters} lists and describes the two main parameters we considered in the analysis: the vicinity radius $r_v$ and the minimum visiting time $t_v^{min}$, both specifying what constitutes a visit of user at a virtual location.
\begin{table}
\small
 \centering
 \begin{tabular}[t]{|c|p{7cm}|}
  \hline
  $r_v$ & \textit{radius of vicinity} of virtual locations: minimum distance (in meter) between users and locations to be considered as visits of the users at locations. \\
  \hline
  $t_v^{min}$ & \textit{minimum visiting time}: minimum time (in seconds) a user has to spend in the vicinity radius of a virtual location to be considered as visit of users at locations. \\
  \hline
 \end{tabular}
 \caption{List of evaluation parameters}
 \label{tab:evaluation-parameters}
\end{table}
\\
\\
\textbf{Virtual locations -- coverage and distribution.}
For our study, we made the simplifying following assumption that the virtual location of a web page is derived from its domain -- that is, all pages with the same domain form the same virtual location. We have collected our current dataset of virtual locations by crawling a major online business directory of the County of Galway\footnote{\url{http://galway.net}}. An a first step, we retrieved the basic information for each entry, i.e. name, address, URL of website (if available), and others. Some entries already provide the corresponding geo coordinates. For all other entries we continued with a second step, namely performing an address resolution using the \textsc{Google Maps} API. As a result, we collected approximately 1,300 entries featuring both a physical and virtual location. This includes 725 locations within the city limits of Galway City. 

\begin{SCfigure*}
  \centering
  \includegraphics[width=0.7\textwidth]{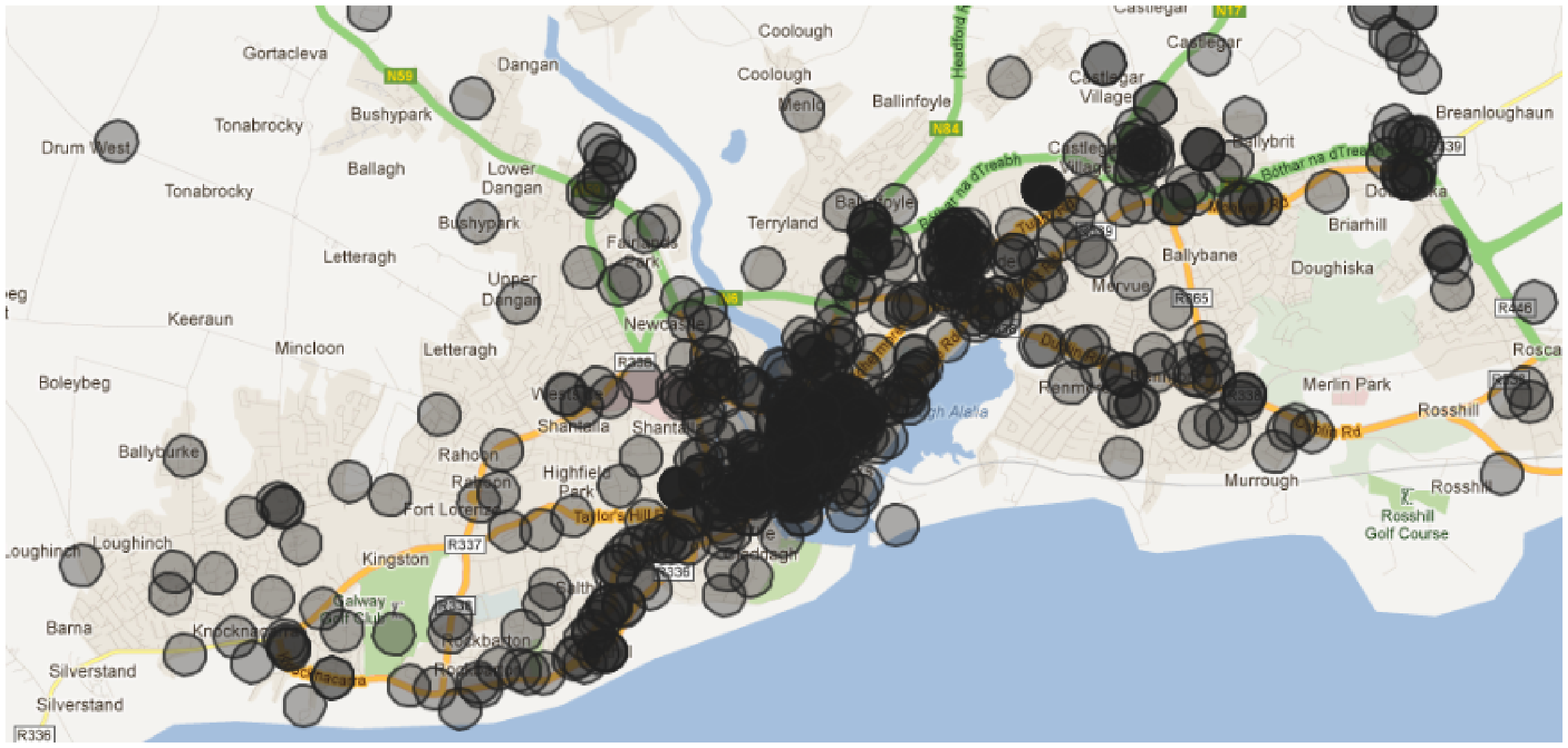}
  \caption{Illustration of coverage ($r = 150\mathrm{m}$).}
  \label{fig:virtual-locations-coverage-map}
\end{SCfigure*}

\begin{figure*}
\parbox{.48\linewidth}{
  \centering
  \includegraphics[width=0.48\textwidth]{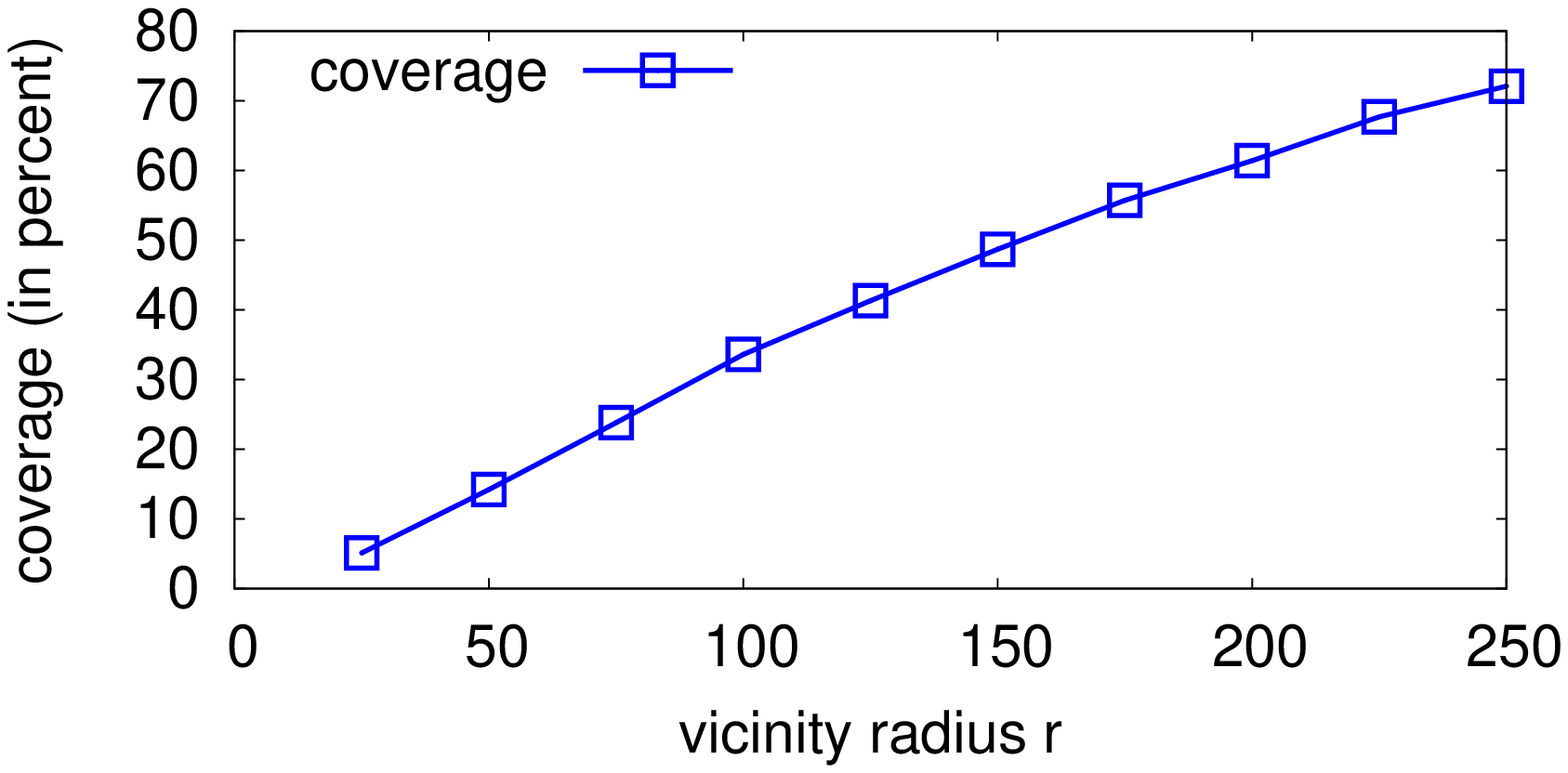}
  \caption{Coverage for different $r_v$.}
  \label{fig:virtual-locations-coverage}
}
\hfill
\parbox{.48\linewidth}{
  \centering
  \includegraphics[width=0.48\textwidth]{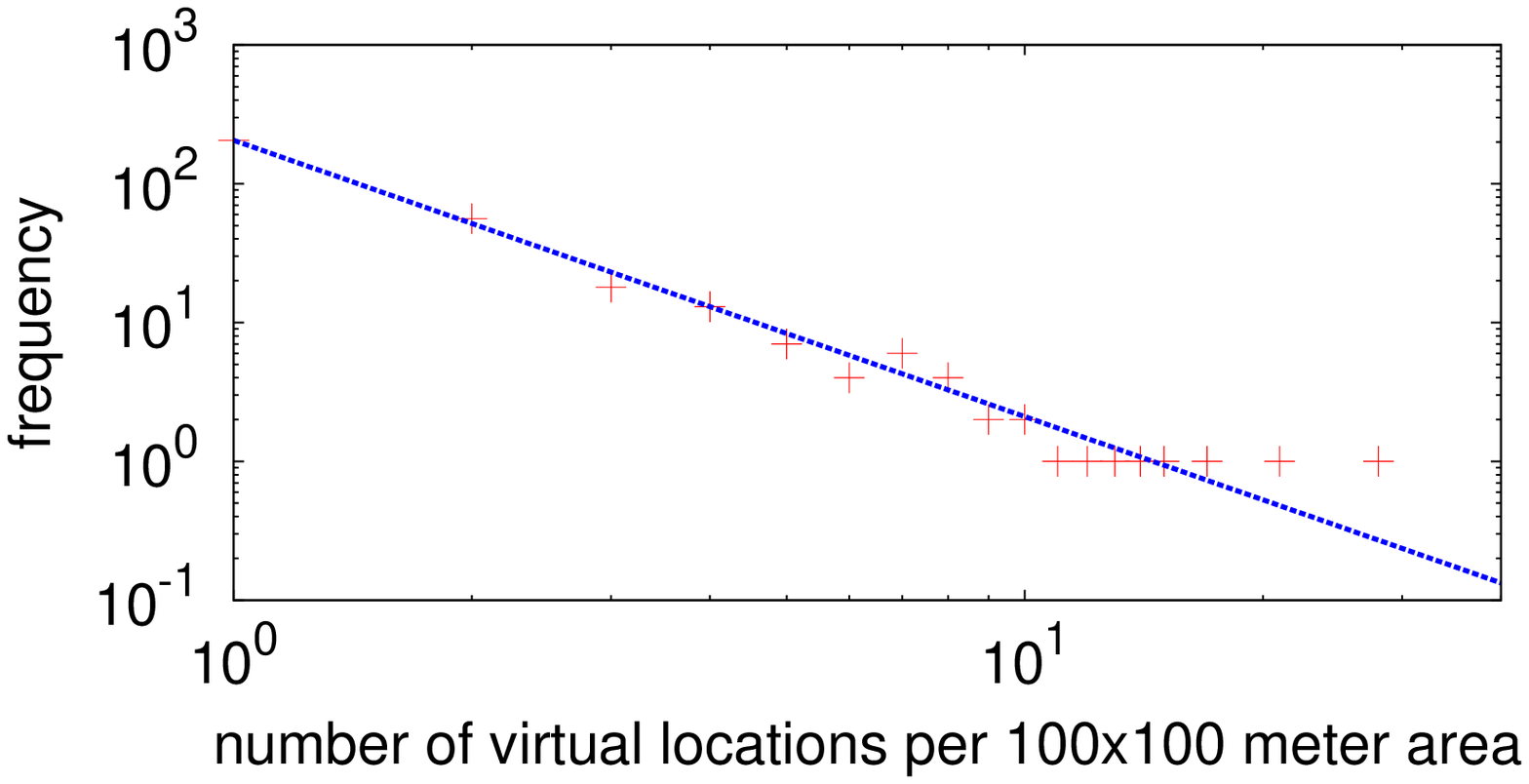}
 \caption{Distribution of non-empty 100$x$100msquares.}
 \label{fig:virtual-locations-distribution}
}
\end{figure*}

We first looked at the coverage, i.e., how much of the area defined by the virtual locations together with their vicinity radiuses overlap with the city of Galway. Figure~\ref{fig:virtual-locations-coverage-map} illustrates the coverage for $r_v = 150\mathrm{m}$. To get more quantitative results, we calculated the coverage in percent; see  Figure~\ref{fig:virtual-locations-coverage}. Naturally, the coverage increases for larger vicinity radiuses, resulting in almost 75\% coverage for $r = 250m$. Figure~\ref{fig:virtual-locations-distribution} shows the distribution all non-empty squares of size 100$x$100 meter. The number of virtual locations per square and their respective frequency show a power-law relationship. That is, while most squares contain only a small set of locations, few squares contain a very large number of virtual locations. 
\\
\\
\textbf{Real-world GPS tracks.}
For our analysis of user movements, we recorded GPS tracks representing two typical uses cases: The~\textit{B2W track} (bike-to-work) track represents one's everyday trip from home to work using a bike through the city center. The distance was 5.4km, the duration was 16:40min ($\mathrm{=1,000sec}$), resulting in an average speed of 19.44km/h. The~\textit{S$^3$ track} (strolling/shopping/sightseeing) is supposed to reflect a typical pastime in the city center, with some (window) shopping, having a snack, some sightseeing, etc. The path was not particularly directed towards specific locations. The overall duration was 2h, and has been recorded in one continuous session and with a resolution of 1 reading per second.

\begin{figure*}
\parbox{.48\linewidth}{
  \centering
  \includegraphics[width=0.48\textwidth]{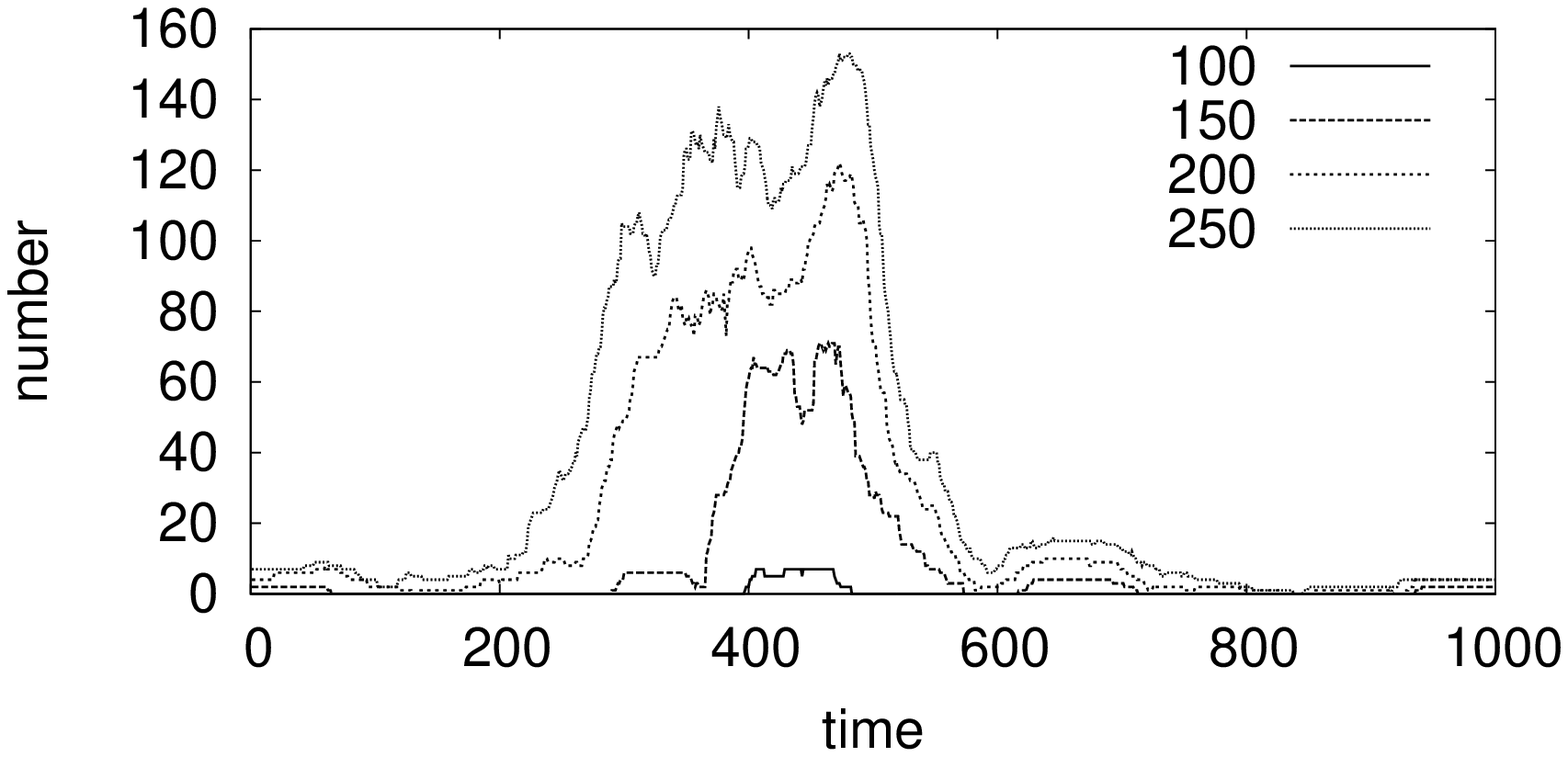}
  \caption{Number of locations visited at the same time: B2W track ($t_v^{min} = 60\mathrm{s}$)}
  \label{fig:parallel-visits-visiting-time-radius-b2w}
}
\hfill
\parbox{.48\linewidth}{
  \centering
  \includegraphics[width=0.48\textwidth]{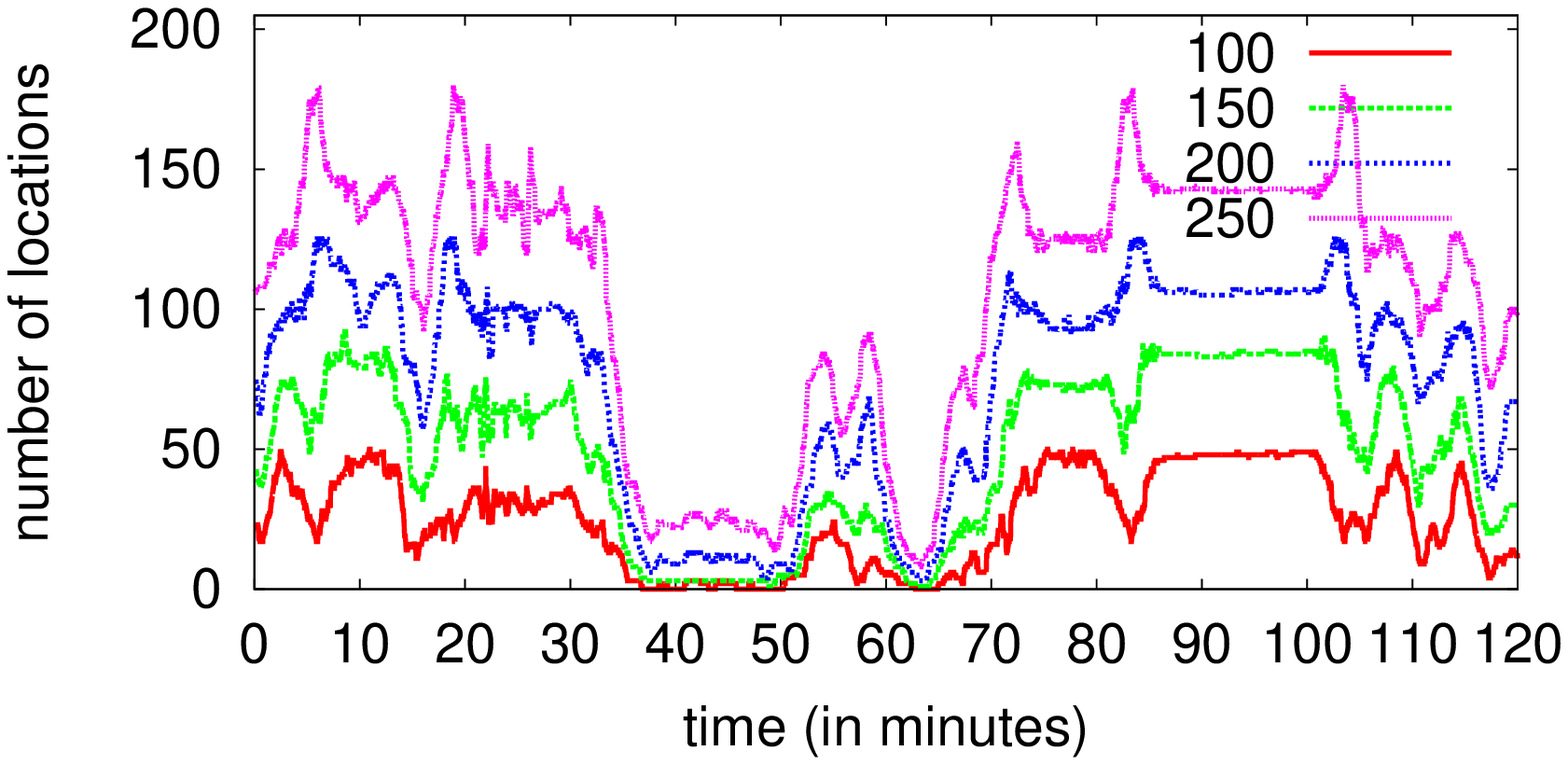}
 \caption{Number of locations visited at the same time: S$^3$ track ($t_v^{min} = 120\mathrm{s}$)}
 \label{fig:parallel-visits-visiting-time-radius-s3}
}
\end{figure*}

We first looked into the number of parallel visits of the user at virtual locations. Figure~\ref{fig:parallel-visits-visiting-time-radius-b2w} and Figure~\ref{fig:parallel-visits-visiting-time-radius-s3} show the results for both tracks and different vicinity radiuses $r_v$. The curve progressions for the B2W track clearly indicate the time interval (200-600), in which the user was close or within the city center where the density of virtual locations is particularly high. In contrast, the curve progressions of the S$^3$ track shows the time (40-50min) when the path led out of the city center. As expected, the number of parallel visits strongly depends on the vicinity radius $r_v$. And lastly, particularly in the city center, the number of parallel visits can be very high. This has direct implications on the implementation of a virtual presence mechanism, since a mobile user is potentially present on many websites in parallel.

\begin{figure*}
\parbox{.48\linewidth}{
  \centering
  \includegraphics[width=0.48\textwidth]{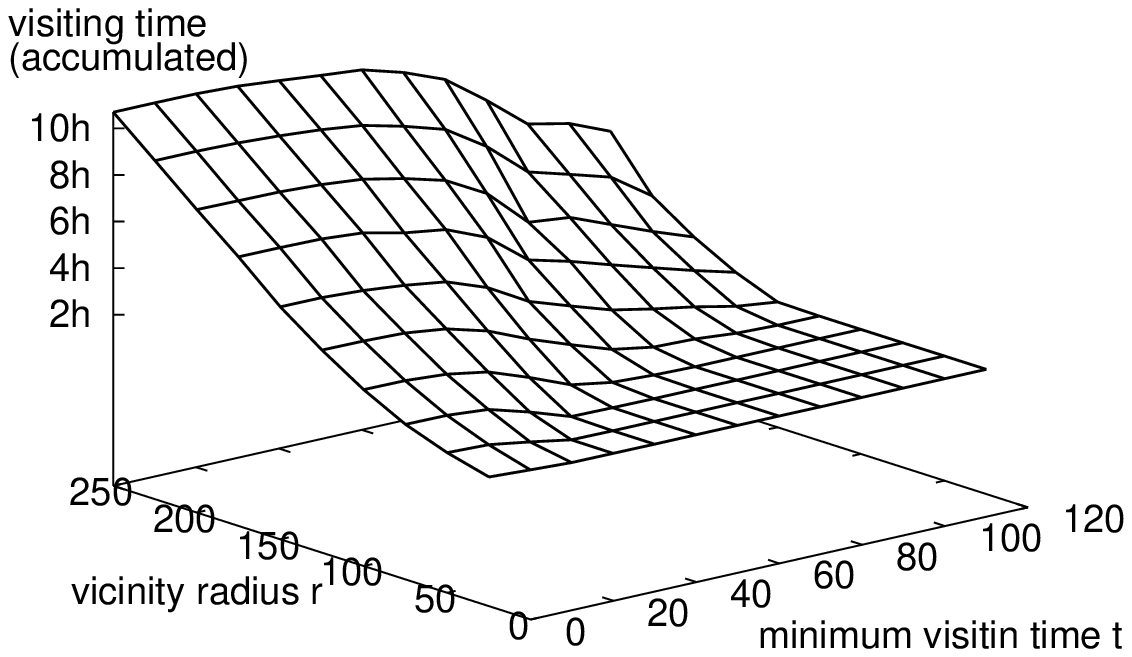}
  \caption{Effect of $r_v$ and $t_v^{min}$ on the accumulated time spent at locations: B2W track}
  \label{fig:cumulative-visiting-time-radius-vs-visittime-b2w}
}
\hfill
\parbox{.48\linewidth}{
  \centering
  \includegraphics[width=0.48\textwidth]{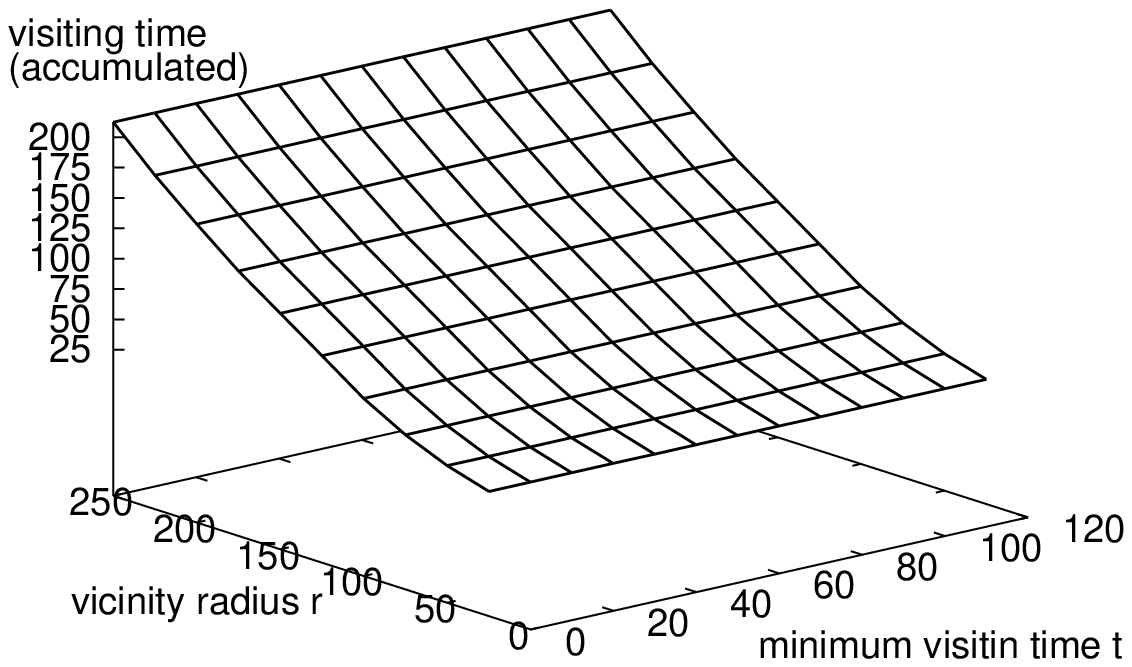}
 \caption{Effect of $r_v$ and $t_v^{min}$ on the accumulated time spent at locations: S$^3$ track}
 \label{fig:cumulative-visiting-time-radius-vs-visittime-s3}
}
\end{figure*}

\begin{figure*}
\parbox{.48\linewidth}{
  \centering
  \includegraphics[width=0.48\textwidth]{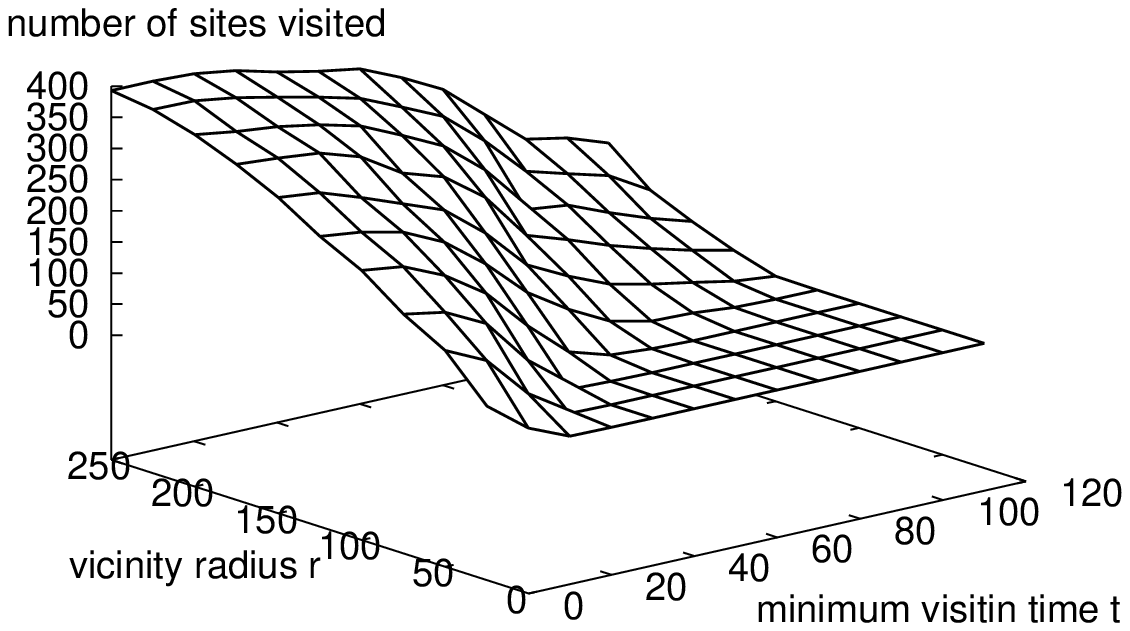}
  \caption{Effect of $r_v$ and $t_v^{min}$ on the overall number of visited locations: B2W track}
  \label{fig:cumulative-num-locations-radius-vs-visittime-b2w}
}
\hfill
\parbox{.48\linewidth}{
  \centering
  \includegraphics[width=0.48\textwidth]{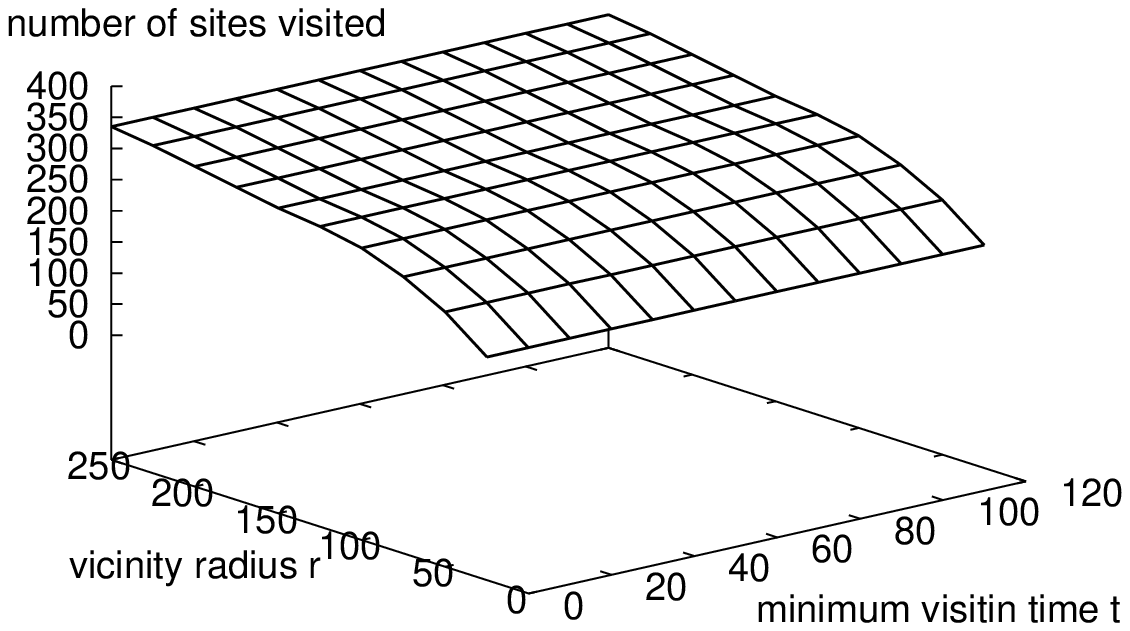}
 \caption{Effect of $r_v$ and $t_v^{min}$ on the overall number of visited locations: S$^3$ track}
 \label{fig:cumulative-num-locations-radius-vs-visittime-s3}
}
\end{figure*}

We then investigated the effect of $r_v$ and $t_v^{min}$ on the accumulated times the user has spent at virtual locations, see Figure~\ref{fig:cumulative-visiting-time-radius-vs-visittime-b2w} and Figure~\ref{fig:cumulative-visiting-time-radius-vs-visittime-s3}, and the overall number of locations the user visited, see Figure~\ref{fig:cumulative-num-locations-radius-vs-visittime-b2w} and Figure~\ref{fig:cumulative-num-locations-radius-vs-visittime-s3}. For the B2W track, both the number of visited locations and the accumulated visiting time increases for larger values of $r_v$ and lower values of $t_v^{min}$. For the S$^3$ track the numbers are almost independent from $t_v^{min}$ due to the much lower traveling speed. Particularly prominent are the high values for the accumulated visiting times. Even for the short B2W track, the accumulated visiting times for the B2W track exceeds 10h for larger values for $r_v$. Given the duration of 2h in the city center, the accumulated visiting time for S$^3$ track is 
significantly higher, up to 200+ hours.

Since the S$^3$ track features a path that crosses and overlaps itself, we looked into the number of revisits. The number is relevant when extending users presence beyond their current locations. Table~\ref{tab:s3-revisit-stats} shows some basic numbers describing the S$^3$ track. Both the number of visited locations and the average duration of visits increase for larger vicinity radiuses $r_v$. The average number of visits of the same locations, however, is relatively stable -- absolute value depends on the chosen path, e.g. how often the path crosses and overlaps itself. The explanation is that increasing values for $r_v$, in general, yield more locations that have been visited multiple times, but also cause the merging of two or more individual visits into one.

\begin{SCtable}
\small
 \centering
 \begin{tabular}[t]{|l|c|c|c|c|}
  \hline
  &	\multicolumn{4}{c|}{vicinity radius $r_v$} \\
  \cline{2-5}
   & 	\textbf{25m} & \textbf{100m} &  \textbf{175m} & \textbf{250m} \\
  \hline\hline
  SUM(locations) & 49  & 235  & 291  & 325 \\
  AVG(duration) & 01:22 & 03:03  & 05:12 & 07:06 \\
  AVG(visits) & 4.86 & 4.15 & 4.83  & 5.49 \\
  \hline
 \end{tabular}
 \caption{S$^3$ track -- basic numbers.}
 \label{tab:s3-revisit-stats}
\end{SCtable}

\section{Conclusions \& On-going Work}
\label{sec:conclusions}
In this paper, we introduced and defined the concept of a virtual presence to support the novel paradigm of ad-hoc socializing between Web users based on shared interests, likings or information needs. We further exploited the fact that many locations feature both a physical and virtual representation to support presence awareness across the boundary between the physical and virtual space. We described a proof-of-concept implementation showcasing the ideas our approach. The analysis of real-world data showed these ideas promise to provide a real practical benefit for both Web and mobile users.

In our on-going work, we head in several directions. 
Firstly, we investigate suitable privacy preservations mechanisms that allow user to specify when they are willing to share their location and to whom. Secondly, we aim to combine our concept of ad-hoc socializing with more traditional channels, particularly social networks, to leverage from existing relationships between users.
Thirdly, we aim to extend our framework to support more semantic connections between physical and virtual locations to enable; for example, a user browsing an article about an event in Galway on a news site is aware of users being actually on-site.
And lastly, we already started the DOBBS\footnote{DOBBS project website: http://dobbs.deri.ie} (DERI Online Browsing Behaviour Study) initiative to create a dataset for the detailed analysis of how users browse the Web. Particularly, DOBBS provides comprehensive information about how long users have visited web pages, which in turn has significant effect on the notion of virtual presence. The datase is public available on the project website.

\vspace{-0.3cm}
\renewcommand{\baselinestretch}{0.97}
\bibliographystyle{abbrv}
\bibliography{literature}

\begin{thebibliography}{10}

\bibitem{Adar08LargeScaleAnalysis}
E.~Adar, J.~Teevan, and S.~T. Dumais.
\newblock {Large Scale Analysis of Web Revisitation Patterns}.
\newblock In {\em Proceedings of the 26th Annual SIGCHI Conference on Human
  Factors in Computing Systems}, CHI '08, pages 1197--1206, New York, NY, USA,
  2008. ACM.

\bibitem{Agichtein06ImprovingWebSearch}
E.~Agichtein, E.~Brill, and S.~Dumais.
\newblock {Improving web Search Ranking by Incorporating User Behavior
  Information}.
\newblock In {\em Proceedings of the 29th Annual International ACM SIGIR
  Conference on Research and Development in Information Retrieval}, SIGIR '06,
  pages 19--26, New York, NY, USA, 2006. ACM.

\bibitem{Amershi08CoSearch}
S.~Amershi and M.~R. Morris.
\newblock {CoSearch: A System for Co-located Collaborative Web Search}.
\newblock In {\em Proceedings of the 26th Annual SIGCHI Conference on Human
  Factors in Computing Systems}, CHI '08, pages 1647--1656, New York, NY, USA,
  2008. ACM.

\bibitem{Antonellis:2008:SQR:1453856.1453903}
I.~Antonellis, H.~G. Molina, and C.~C. Chang.
\newblock {SimRank++: Query Rewriting Through Link Analysis of the Click
  Graph}.
\newblock {\em Proc. VLDB Endow.}, 1(1):408--421, Aug. 2008.

\bibitem{Bohunsky10VisualStructureBased}
P.~Bohunsky and W.~Gatterbauer.
\newblock {Visual Structure-based Web Page Clustering and Retrieval}.
\newblock In {\em Proceedings of the 19th International Conference on World
  Wide Web}, WWW '10, pages 1067--1068, New York, NY, USA, 2010. ACM.

\bibitem{Broder97SyntacticClustering}
A.~Z. Broder, S.~C. Glassman, M.~S. Manasse, and G.~Zweig.
\newblock {Syntactic clustering of the Web}.
\newblock In {\em Selected Papers From the 6th International Conference on
  World Wide Web}, pages 1157--1166, Essex, UK, 1997. Elsevier Science
  Publishers Ltd.

\bibitem{Chakrabarti98Automatic}
S.~Chakrabarti, B.~Dom, P.~Raghavan, S.~Rajagopalan, D.~Gibson, and
  J.~Kleinberg.
\newblock {Automatic Resource Compilation by Analyzing Hyperlink Structure and
  Associated Text}.
\newblock In {\em Proceedings of the 7th International Conference on World Wide
  Web}, WWW'07, pages 65--74, Amsterdam, The Netherlands, The Netherlands,
  1998. Elsevier Science Publishers B. V.

\bibitem{DaZhen07NearReplicas}
W.~Da-Zhen and C.~Yu-Hui.
\newblock {Near-Replicas of Web Pages Detection Efficient Algorithm Based on
  Single MD5 Fingerprint}.
\newblock In {\em Proceedings of the 8th Conference on 8th WSEAS International
  Conference on Automation and Information - Volume 8}, ICAI'07, pages
  318--320, Stevens Point, Wisconsin, USA, 2007. World Scientific and
  Engineering Academy and Society (WSEAS).

\bibitem{Dubroy10AStudyOfTabbed}
P.~Dubroy and R.~Balakrishnan.
\newblock {A Study of Tabbed Browsing Among Mozilla Firefox Users}.
\newblock In {\em Proceedings of the 28th International Conference on Human
  Factors in Computing Systems}, CHI '10, pages 673--682, New York, NY, USA,
  2010. ACM.

\bibitem{Fu06DetectingFishing}
A.~Y. Fu, L.~Wenyin, and X.~Deng.
\newblock {Detecting Phishing Web Pages with Visual Similarity Assessment Based
  on Earth Mover's Distance (EMD)}.
\newblock {\em IEEE Transactions on Dependable and Secure Computing},
  3(4):301--311, Oct. 2006.

\bibitem{Goel12WhoDoesWhat}
S.~Goel, J.~M. Hofman, and M.~I. Sirer.
\newblock {Who Does What on the Web: A Large-scale Study of Browsing Behavior}.
\newblock In {\em Proceedings of the 6th International Conference on Weblogs
  and Social Media}, ICWSM'12.

\bibitem{Gueting05Moving}
R.~G{\"u}ting and M.~Schneider.
\newblock {\em {Moving Objects Databases}}.
\newblock Morgan Kaufmann Series in Data Management Systems. Morgan Kaufmann,
  2005.

\bibitem{Jeh02SimRank}
G.~Jeh and J.~Widom.
\newblock {SimRank: A Measure of Structural-context Similarity}.
\newblock In {\em Proceedings of the eighth ACM SIGKDD International Conference
  on Knowledge Discovery and Data Mining}, KDD '02, pages 538--543, New York,
  NY, USA, 2002. ACM.

\bibitem{Kleinberg99HITS}
J.~M. Kleinberg.
\newblock {Authoritative Sources in a Hyperlinked Environment}.
\newblock {\em Journal of the ACM}, 46(5):604--632, Sept. 1999.

\bibitem{Kritikopoulos07WordRank}
A.~Kritikopoulos, M.~Sideri, and I.~Varlamis.
\newblock {WordRank: A Method for Ranking Web Pages Based on Content
  Similarity}.
\newblock In {\em Proceedings of the 24th British National Conference on
  Databases}, BNCOD '07, pages 92--100, Washington, DC, USA, 2007. IEEE
  Computer Society.

\bibitem{Leshed08CoScripter}
G.~Leshed, E.~M. Haber, T.~Matthews, and T.~Lau.
\newblock {CoScripter: Automating \& Sharing How-to Knowledge in the
  Enterprise}.
\newblock CHI '08, pages 1719--1728, New York, NY, USA, 2008. ACM.

\bibitem{Li05WebDataExtraction}
Z.~Li, W.~K. Ng, and A.~Sun.
\newblock {Web Data Extraction Based on Structural Similarity}.
\newblock {\em Knowledge and Information Systems}, 8(4):438--461, Nov. 2005.

\bibitem{Lin09MatchSim}
Z.~Lin, M.~R. Lyu, and I.~King.
\newblock {MatchSim: A Novel Neighbor-based Similarity Measure with Maximum
  Neighborhood Matching}.
\newblock In {\em Proceedings of the 18th ACM Conference on Information and
  Knowledge Management}, CIKM '09, pages 1613--1616, New York, NY, USA, 2009.
  ACM.

\bibitem{Meiss09WhatsInASession}
M.~Meiss, J.~Duncan, B.~Gon\c{c}alves, J.~J. Ramasco, and F.~Menczer.
\newblock {What's in a Session: Tracking Individual Behavior on the Web}.
\newblock In {\em Proceedings of the 20th ACM Conference on Hypertext and
  Hypermedia}, HT '09, pages 173--182, New York, NY, USA, 2009. ACM.

\bibitem{Morris08Survey}
M.~R. Morris.
\newblock {A Survey of Collaborative Web Search Practices}.
\newblock CHI '08, pages 1657--1660, New York, NY, USA, 2008. ACM.

\bibitem{Morris07SearchTogether}
M.~R. Morris and E.~Horvitz.
\newblock {SearchTogether: An Interface for Collaborative Web Search}.
\newblock In {\em Proceedings of the 20th Annual ACM Symposium on User
  Interface Software and Technology}, UIST '07, pages 3--12, New York, NY, USA,
  2007. ACM.

\bibitem{Morris06TeamSearch}
M.~R. Morris, A.~Paepcke, and T.~Winograd.
\newblock {TeamSearch: Comparing Techniques for Co-Present Collaborative Search
  of Digital Media}.
\newblock In {\em Proceedings of the First IEEE International Workshop on
  Horizontal Interactive Human-Computer Systems}, TABLETOP '06, pages 97--104,
  Washington, DC, USA, 2006. IEEE Computer Society.

\bibitem{Page98PageRank}
L.~Page, S.~Brin, R.~Motwani, and T.~Winograd.
\newblock {The PageRank Citation Ranking: Bringing Order to the Web}.
\newblock {\em World Wide Web Internet And Web Information Systems},
  (1999-66):1--17, 1998.

\bibitem{Shivakumar98FindingNearReplicas}
N.~Shivakumar and H.~Garcia-Molina.
\newblock {Finding Near-Replicas of Documents on the Web}.
\newblock In {\em Selected Papers From the International Workshop on The World
  Wide Web and Databases}, WebDB'98, pages 204--212, London, UK, 1999.
  Springer-Verlag.

\bibitem{vdw11COBS}
C.~von~der Weth and A.~Datta.
\newblock {COBS: Realizing Decentralized Infrastructure for Collaborative
  Browsing and Search}.
\newblock In {\em 2011 IEEE International Conference on Advanced Information
  Networking and Applications}, pages 617--624. IEEE, Mar. 2011.

\bibitem{Wedig06LargeScaleAnalysis}
S.~Wedig and O.~Madani.
\newblock {A Large-scale Analysis of Query Logs for Assessing Personalization
  Opportunities}.
\newblock In {\em Proceedings of the 12th ACM SIGKDD International Conference
  on Knowledge Discovery and Data Mining}, KDD '06, pages 742--747, New York,
  NY, USA, 2006. ACM.

\bibitem{Xue10UserNavigation}
L.~Xue, M.~Chen, Y.~Xiong, and Y.~Zhu.
\newblock {User Navigation Behavior Mining Using Multiple Data Domain
  Description}.
\newblock In {\em Proceedings of the 2010 IEEE/WIC/ACM International Conference
  on Web Intelligence and Intelligent Agent Technology - Volume 03}, WI-IAT
  '10, pages 132--135, Washington, DC, USA, 2010. IEEE Computer Society.

\bibitem{Zhang11MeasuringWebPage}
H.~Zhang and S.~Zhao.
\newblock {Measuring Web Page Revisitation in Tabbed Browsing}.
\newblock In {\em Proceedings of the 2011 Annual Conference on Human Factors in
  Computing Systems}, CHI '11, pages 1831--1834, New York, NY, USA, 2011. ACM.

\end{thebibliography}

\end{document}